\documentclass[12pt]{article}
\usepackage{epsfig, epsf, graphicx}
\usepackage{pstricks, pst-node, psfrag}
\usepackage{amssymb,amsmath,bm}
\usepackage{verbatim,enumerate}
\usepackage{diagbox}
\usepackage{rotating, lscape}
\usepackage{setspace}
\usepackage{paralist}
\usepackage{apacite}
\usepackage{natbib}
\usepackage{color, colortbl}
\usepackage{tabu}
\usepackage{subcaption}
\usepackage{xcolor}
\usepackage{contour}
\usepackage{float}
\usepackage{multirow}

\definecolor{seablue}{RGB}{0, 103, 225}

\newcommand{\rev}[1]{\textcolor{black}{#1}}

\usepackage[framemethod=default]{mdframed}
\usepackage{showexpl}

\mdfdefinestyle{exampledefault}{%
linecolor=black,linewidth=1.5pt,
middlelinewidth=3pt,rightline=true,innerleftmargin=10,innerrightmargin=10}
\usepackage{tikz}
\usepackage{multirow}
\usepackage{bbm}
\usetikzlibrary{arrows,calc,tikzmark,shapes,fit,positioning}
\tikzset{box/.style={draw, rectangle, thick, text centered, minimum height=3em}}
  \tikzset{line/.style={draw, thick, -latex'}}
\usepackage{hyperref}
\usepackage{booktabs}
\usepackage{float}
\usepackage{multicol,listings,siunitx}
\usepackage{caption}
\usepackage{gensymb}

\usepackage{adjustbox}

\def\boxit#1{\vbox{\hrule\hbox{\vrule\kern6pt
          \vbox{\kern6pt#1\kern6pt}\kern6pt\vrule}\hrule}}

\def\bse{\begin{eqnarray*}}
\def\ese{\end{eqnarray*}}
\def\be{\begin{eqnarray}}
\def\ee{\end{eqnarray}}
\def\bq{\begin{equation}}
\def\eq{\end{equation}}
\def\bse{\begin{eqnarray*}}
\def\ese{\end{eqnarray*}}

\definecolor{seagreen}{rgb}{0.18,0.55,0.34}
\definecolor{lawngreen}{rgb}{0.49,0.99,0}
\definecolor{lightsalmon}{rgb}{1,0.63,0.48}
\definecolor{lightyellow}{rgb}{0.99,0.906,0.429}

\newcommand{\bx}{\mathbf{x}}

\newtheorem{prop}{Proposition}

\DeclareFontFamily{OT1}{pzc}{}
\DeclareFontShape{OT1}{pzc}{m}{it}{<-> s * [1.10] pzcmi7t}{}
\DeclareMathAlphabet{\mathpzc}{OT1}{pzc}{m}{it}

\begin{document}

\thispagestyle{empty} \baselineskip=28pt \vskip 5mm
\begin{center} {\LARGE{\bf 3D Bivariate Spatial Modelling of Argo Ocean Temperature and Salinity}}
	
\end{center}

\baselineskip=12pt \vskip 10mm

\begin{center}\large
Mary Lai O. Salva\~{n}a\footnote[1]{\baselineskip=10pt Department of Statistics, University of Connecticut, Email: marylai.salvana@uconn.edu},  Jian Cao\footnote[2]{\baselineskip=10pt Department of Mathematics, University of Houston, Email: jcao21@central.uh.edu }, and Mikyoung Jun\footnote[3]{Corresponding author. \baselineskip=10pt Department of Mathematics, University of Houston, Email: mjun@central.uh.edu\\
Mikyoung Jun acknowledges support by NSF DMS-1925119, DMS-2105847, and DMS-2413042. The authors also acknowledge helpful discussions with Mikael Kuusela on Argo data.}
\end{center}

\baselineskip=17pt \vskip 10mm \centerline{\today} \vskip 15mm


\begin{center}
{\large{\bf Abstract}}
\end{center} 
Variables within the global oceans can reveal the impacts of a warming climate, as the oceans absorb huge amounts of solar energy. Understanding the joint spatial distribution of key ocean variables is therefore essential. In this paper, we investigate the spatial dependence structure between ocean temperature and salinity using Argo observations and construct a bivariate spatial model covering from the surface through the ocean interior. We develop a flexible class of multivariate nonstationary covariance models defined in 3-dimensional (3D) space (longitude $\times$ latitude $\times$ depth) that allow the variances and correlations to vary with depth, capturing the ocean’s vertical structure. These models describe the joint spatial distribution of the two variables while incorporating the underlying vertical structure of the ocean. We apply this framework to Argo temperature and salinity data and address the computational challenges of large data volumes through the Vecchia approximation. Our results show that the proposed bivariate covariance model effectively represents the complex vertical cross-covariance structure of the processes and their first- and second-order differences, whereas classical bivariate models, including the bivariate Mat\'{e}rn, poorly fit the empirical cross-covariance structure.

\par\vfill\noindent
{\bf Some key words:} 3D covariance functions; Argo; cross-covariance function; nonstationary; salinity; spatial; temperature.

\clearpage\pagebreak\newpage \pagenumbering{arabic}
\baselineskip=26pt


\section{Introduction} \label{sec:intro}

\begin{figure}[hbt!]
\centering
 \includegraphics[scale=0.25]{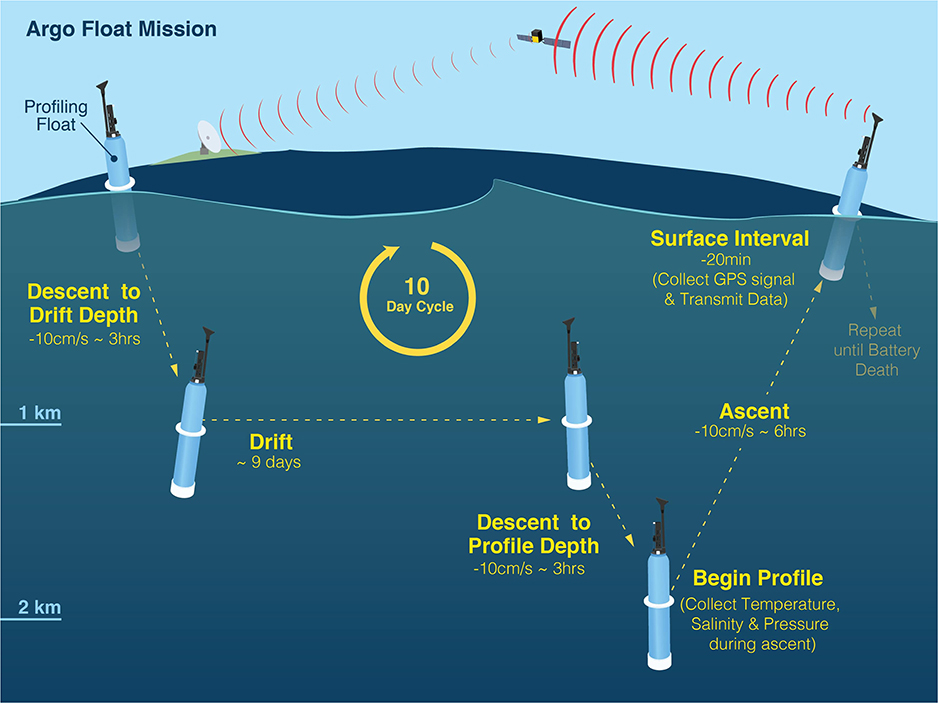}
 \caption{Standard Argo ``park-and-profile'' mission (Source: \cite{wong2020argo})}
 \label{fig:argo_park_and_profile}
\end{figure}

The international program named Array for Real-time Geostrophic Oceanography (ARGO) was launched in the early 2000s in response to the need for a global ocean observing system \citep{wong2020argo, johnson2022argo}. Since then, the program has deployed a worldwide network of 4,000 free-drifting profiling floats that sample the upper 2,000 meters of the oceans. Each float completes a ten-day ``park-and-profile'' mission: it descends from the surface to a drift depth of 1,000 meters, parks there for nine days, then descends to 2,000 meters and measures temperature ($^{\circ}$C), salinity (in practical salinity unit or PSU), and pressure (in decibar or dbar) while it ascends to the surface. Upon surfacing, the float transmits the collected data via satellite. The full cycle is shown in Figure~\ref{fig:argo_park_and_profile}. All measurements are made available within 24 hours as Argo data products (\textcolor{blue}{\url{https://argo.ucsd.edu}}), providing both opportunities and challenges for analyzing ocean processes.

The core Argo data products are temperature and salinity--the two fundamental variables from which many other oceanographic variables can be derived, including freezing point, electrical conductivity, and viscosity \citep{pawlowicz2013key}, ocean heat content and potential density \citep{yarger2022functional}, and tropical cyclone-induced ocean thermal response \citep{hu2024spatiotemporal}. Temperature and salinity are central to understanding the oceans' physical properties and dynamics: their distributions influence ocean circulation \citep{chen2022effects, gangopadhyay2022introduction}, shape climate processes \citep{olson2022effect}, and alter biogeochemistry \citep{galan2021argo, ding2022effects}. They also support diverse applications, including hydroacoustics \citep{jana2022sound}, offshore wind farm design \citep{escobar2016influence}, and assessments of environmental stressors in marine and freshwater systems \citep{walker2020warmer}.

In most studies, including those noted above, Argo temperature and salinity are modeled separately. Here, we jointly model them to account for spatial cross-correlation. The proposed bivariate model in 3D space offers sufficient expressiveness across depth to capture complex oceanographic structures. Because ocean density varies continuously with depth, the water column becomes vertically stratified, reflecting the mixing of distinct water masses—for example, cold, fresh subarctic waters and warm, salty subtropical waters \citep{sambe2022unsupervised}. Since temperature and salinity characterize these water masses, stratification directly shapes their spatial distributions. Bivariate spatial models that incorporate vertical structure can therefore provide a more faithful representation of the underlying processes. 

\subsection{Spatial covariance models in 2D and 3D space}

\begin{figure}[bpht!]
	\centering
	\includegraphics[scale=0.15]{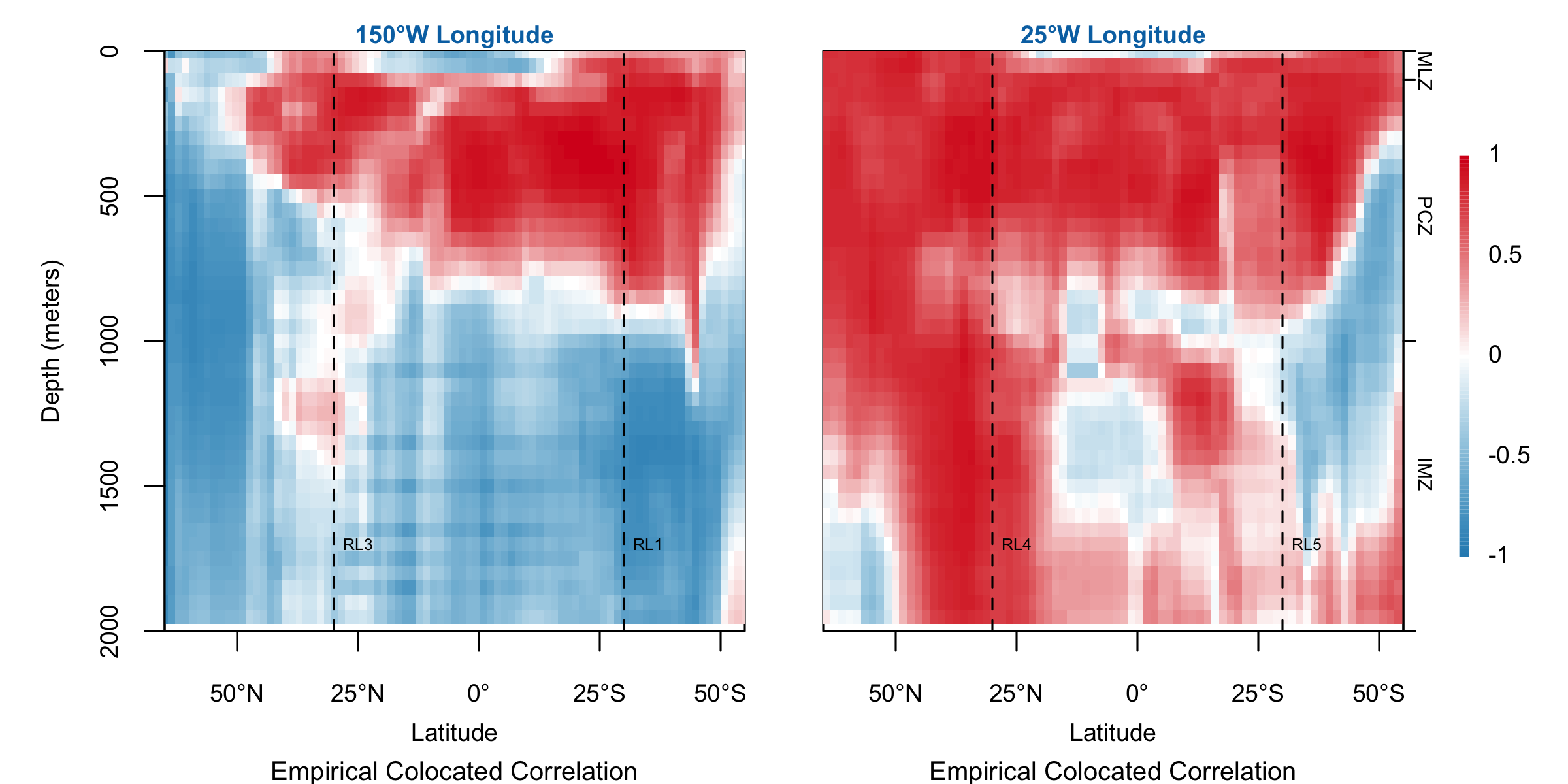}
    \caption{Empirical colocated correlation of temperature and salinity residuals from January to March in 2016.}
	\label{fig:empirical_colocated_correlation_vertical}
\end{figure}

Spatial variation—and in particular the cross-correlation between Argo temperature and salinity—exhibits complex structure both horizontally and vertically. 
Figure~\ref{fig:empirical_colocated_correlation_vertical} shows the empirical colocated correlation between the two variables across latitude and depth at two longitudinal levels. Here, we use temperature and salinity residuals from \cite{yarger2022functional}. See Section 2.1 for more details on the residuals. The correlation of the residuals of the two variables spans nearly the full range $(-1, 1)$, with vertical patterns that differ markedly by latitude and longitude. These features highlight the need for flexible cross-covariance functions developed in 3D space capable of capturing the pronounced nonstationarity along the depth dimension. 

Traditionally, spatial domains are modeled in two dimensions, and the literature provides a wide range of 2D cross-covariance functions. Among the most widely used is the parsimonious Mat\'{e}rn cross-covariance function of \cite{gneiting2010matern}, which forms the basis for many subsequent developments. Its general form is
\begin{equation}\label{eqn:multivariate_matern}
C_{ij}(\mathbf{s}_1, \mathbf{s}_2) = \frac{\rho_{ij} \sigma_{i} \sigma_{j}}{2^{\nu_{ij} - 1} \Gamma \left( \nu_{ij} \right)} \mathcal{M}_{\nu_{ij}} \left\{h (\mathbf{s}_1, \mathbf{s}_2)\right\}, \quad i,j = 1, \ldots, q,
\end{equation}
where $h(\mathbf{s}_1, \mathbf{s}_2)$ is a distance function on $\mathbb{R}^{d}$, $\mathcal{M}_{\nu}(x) = x^{\nu}\mathcal{K}_{\nu}(x)$, $\mathcal{K}_{\nu}$ is the modified Bessel function of the second kind, $\Gamma$ is the gamma function, and $q$ is the number of variables. The model includes marginal variance ($\sigma_i^2>0$) and smoothness ($\nu_{ii}>0$) parameters for $i=j$, and colocated correlation ($\rho_{ij}$) and cross-smoothness ($\nu_{ij}>0$) parameters for $i\neq j$, where $\nu_{ij} = \frac{1}{2} (\nu_{ii} + \nu_{jj})$ and 
\begin{equation} \label{eqn:colocated_correlation}
\rho_{ij} = \beta_{ij} \frac{\Gamma(\nu_{ii} + \frac{d}{2})^{1/2}}{\Gamma(\nu_{ii})^{1/2}} \frac{\Gamma(\nu_{jj} + \frac{d}{2})^{1/2}}{\Gamma(\nu_{jj})^{1/2}} \frac{\Gamma(\nu_{ij} ) }{\Gamma( \nu_{ij} + \frac{d}{2})},
\end{equation}
for any $\sigma_{i}^2, \nu_{ii} > 0$, $d \in \mathbb{N}$. Here, $(\beta_{ij})_{i, j = 1}^{q}$ is a symmetric and positive definite correlation matrix. 

Certain choices of the distance functions $h (\mathbf{s}_1, \mathbf{s}_2)$ may render the parsimonious Mat\'{e}rn cross-covariance non-positive definite. For 3D spatial processes, a suitable distance is:
\begin{eqnarray}
\label{eqn:distance_formula}
h(L_1, L_2, l_1 - l_2, p_1 - p_2) & = & {a_{\text{h}}^2 \text{ch}^2(L_1, L_2, l_1 - l_2) + a_{\text{v}}^2 (p_1 - p_2)^2},
\end{eqnarray}
where $a_{\text{h}}$ and $a_{\text{v}}$ are horizontal and vertical scale parameters, respectively. Horizontal separation is measured by the chordal distance between latitude–longitude pairs,
\begin{eqnarray}
\text{ch}(L_1, L_2, l_1 - l_2) = 2 R \left\{ \sin^2 \left( \frac{L_1 - L_2}{2} \right) + \cos L_1 \cos L_2 \sin^2 \left( \frac{l_1 - l_2}{2} \right) \right\}^{1/2},
\end{eqnarray}
and vertical separation is measured as the Euclidean distance in depth. Here, $R$ is Earth’s radius, taken as 6371 km. For completeness, we note that chordal distances are often used to extend valid covariance functions from flat to spherical domains \citep{jun2007approach}. Further developments for spherical settings include \cite{alegria2020cross} and \cite{cao2022locally}; see \cite{sphere_review} for a comprehensive review of covariance models on spheres. 

Even if one can define a covariance model in 3D with the distance as defined in \eqref{eqn:distance_formula}, the Mat\'{e}rn cross-covariance function in (\ref{eqn:multivariate_matern}) remains stationary: the variances $\sigma_{i}^2$, colocated correlations $\rho_{ij}$, and scale parameters $a_{\text{h}}$ and $a_{\text{v}}$ are constant in space. To relax this assumption in 2D, \cite{kleiber2012nonstationary} proposed a nonstationary extension, allowing variances and scales to vary spatially:
\begin{equation}\label{eqn:kleiber_nychka_nonstationary}
C_{ij}(\mathbf{s}_1, \mathbf{s}_2) = \rho_{ij} \frac{\sigma_i (\mathbf{s}_1)  \sigma_j (\mathbf{s}_2)}{|\mathbf{D}_{ij}(\mathbf{s}_1, \mathbf{s}_2)|^{1/2}} \mathcal{M}_{\nu_{ij}} \left\{ \left( (\mathbf{s}_1 - \mathbf{s}_2)^{\top}\mathbf{D}_{ij}(\mathbf{s}_1, \mathbf{s}_2)^{-1}(\mathbf{s}_1 - \mathbf{s}_2) \right)^{1/2} \right\}, 
\end{equation}
where $\mathbf{D}_{ij}(\mathbf{s}_1, \mathbf{s}_2) = \frac{1}{2}\left\{\mathbf{D}_i(\mathbf{s}_1)+\mathbf{D}_j(\mathbf{s}_2 )\right\}$ is a $d\times d$ positive definite kernel matrix that controls spatially varying anisotropy, for $i,j = 1, \ldots, q$. Despite this innovation, extending the Mat\'{e}rn model to include nonstationary colocated correlations $\rho_{ij}(\mathbf{s})$ remains challenging. One proposed remedy is to replace the constants $\sigma_i^2$ and $\beta_{ij}$ with kernel-smoothed nonparametric estimators \citep{kleiber2012nonstationary}:
\begin{equation} \label{eqn:nonparametric_variance}
\begin{gathered}
    \hat{\sigma}_{i}^2(\mathbf{s}) = \frac{\sum_{\tilde{\mathbf{s}} \in \tilde{\mathcal{D}}} K_{\lambda_{h}, \lambda_{v}}(\mathbf{s}, \tilde{\mathbf{s}}) Z_i^2 (\tilde{\mathbf{s}})}{ \sum_{\tilde{\mathbf{s}} \in \tilde{\mathcal{D}}} K_{\lambda_{h}, \lambda_{v}}(\mathbf{s}, \tilde{\mathbf{s}})} \quad \text{ and } \\ 
     \hat{\beta}_{ij}(\mathbf{s}_1, \mathbf{s}_2) = \frac{\sum_{\tilde{\mathbf{s}} \in \tilde{\mathcal{D}}} K_{\lambda_{h}, \lambda_{v}}(\mathbf{s}_1, \tilde{\mathbf{s}})^{1/2} K_{\lambda_{h}, \lambda_{v}}(\mathbf{s}_2, \tilde{\mathbf{s}})^{1/2} Z_i (\tilde{\mathbf{s}}) Z_j (\tilde{\mathbf{s}})}{\hat{\sigma}_{i}(\mathbf{s}_1) \hat{\sigma}_{j}(\mathbf{s}_2) \{\sum_{\tilde{\mathbf{s}} \in \tilde{\mathcal{D}}} K_{\lambda_{h}, \lambda_{v}}(\mathbf{s}_1, \tilde{\mathbf{s}})\}^{1/2} \{\sum_{\tilde{\mathbf{s}} \in \tilde{\mathcal{D}}} K_{\lambda_{h}, \lambda_{v}}(\mathbf{s}_2, \tilde{\mathbf{s}})\}^{1/2}},
     \end{gathered}
\end{equation}
where $K_{\lambda_{h}, \lambda_{v}}(\mathbf{s}, \tilde{\mathbf{s}}) = \exp[-\{\text{ch}^2(L_1, L_2, l_1 - l_2) / \lambda_{h} + (p_1 - p_2)^2 / \lambda_{v} \}]$ is a nonnegative kernel with horizontal and vertical bandwidths $\lambda_{h}$ and $\lambda_{v}$, respectively, and $Z_i(\mathbf{s})$ is the observed residual field. However, this approach does not guarantee that the resulting cross-covariance matrix is positive definite; this must be enforced post hoc by adjusting negative eigenvalues. \cite{kleiber2013spatially} developed sufficient conditions for positive definiteness regarding $\beta_{ij}(\mathbf{s}_1,\mathbf{s}_2)$, but their approach requires a nugget effect and covariate regression. In contrast, we develop flexible parametric cross-covariance models in 3D using differential operators, with nonstationarity introduced primarily along the depth dimension through basis expansions. The proposed models retain closed-form expressions, enabling straightforward likelihood-based inference without requiring covariates (which may not always be available). As shown in our real-data application, these models capture not only the dependence in the original processes but also in their differenced versions, which has important implications for parameter estimation and Kriging \citep{Stein1999}. The rest of the paper is structured as follows: Section~\ref{sec:background} describes the data and presents various empirical results, along with a review of differential operator methods. Section~\ref{sec:methodology} introduces the 3D bivariate spatial models and discusses computational aspects. Section~\ref{sec:application} reports model fitting, prediction, and diagnostics. Section~\ref{sec:discussion} concludes the paper.

\section{Background} \label{sec:background}

We first review several spatial methods applied to Argo data, followed by a brief overview of differential operator approaches in 2D that motivate our proposed 3D model.

\subsection{Spatial Interpolation of Argo Data}

Spatially continuous maps of temperature and salinity measurements are essential to multidisciplinary scientific research. However, despite the planetary scale and subsurface reach of the Argo profiling network, not all locations are sampled. Since other ocean variables depend heavily on temperature and salinity, efforts are focused on obtaining the best interpolated maps of these two variables. Several institutions have produced high-resolution, gridded global temperature and salinity datasets from the sparse measurements recorded by the Argo floats using various interpolation techniques \citep{liu2020global}. Examples include the EN4 dataset from the U.K.\ Met Office \citep{good2013en4}, the Monthly Objective Analysis using Argo data (MOAA) from JAMSTEC \citep{hosoda2008monthly}, and the BOA-Argo dataset, which applies objective interpolation via the Barnes successive correction method \citep{li2017development}. 

One widely used Argo product is the Roemmich–Gilson Argo climatology from the Scripps Institution of Oceanography \citep{roemmich2009}, which employs weighted local regression. Let $\mathbf{s} = (L, l, p)$ denote spatial location, where $L$ is latitude, $l$ is longitude, and $p$ is pressure (commonly treated as depth in the upper ocean). Let $t$ denote time in yeardays (days since January 1 of a given year). For a reference location $\mathbf{s}_0 = (L_0, l_0, p_0)^{\top}$, \cite{roemmich2009} fit the mean function:
\begin{multline}
\mu_{\mathbf{s}_0} (\mathbf{s}, t) = \beta_0 + \beta_1 (L - L_0) + \beta_2 (l - l_0) + \beta_3 (L - L_0)^2 + \beta_4 (l - l_0)^2 + \beta_5 (p - p_0)  \\
+ \beta_6 (p - p_0)^2+ \sum_{k = 1}^6 \gamma_k \sin \left( 2 \pi k \frac{t}{365.25} \right)+ \sum_{k = 1}^6 \delta_k \cos \left( 2 \pi k \frac{t}{365.25} \right)
\label{eqn:roemmich_mean}
\end{multline}
using observations from the nearest horizontal neighbors, pressure levels and each of the
12 calendar months, drawn from the entire Argo record. Here, $\beta_0$ and $\beta_k$, $\gamma_k$, and $\delta_k$, $k = 1, \ldots, 6$, are scalar coefficients. The residuals were modeled using the spatial covariance function:
\begin{align}
C (L_1, L_2; l_1, l_2) \propto 0.77 \exp \left[ - \left\{ \frac{h_{\text{RG}}(L_1, L_2; l_1, l_2)}{140 \text{ km}} \right\}^2 \right] + 0.23 \exp \left\{ - \frac{h_{\text{RG}}(L_1, L_2; l_1, l_2)}{1111 \text{ km}} \right\},
\end{align}
where the distance
\begin{equation*}
h_{\text{RG}}(L_1, L_2; l_1, l_2) = \sqrt{(L_1 - L_2)^2 + (l_1 - l_2)^2 + \text{penalty}(L_1, L_2; l_1, l_2)^2}.
\end{equation*}
includes a penalty to reduce correlation across strong physical boundaries. \cite{kuusela2018locally} improved the residual modelling with an anisotropic space-time exponential covariance function
 \begin{equation}
C (L_1, L_2; l_1, l_2; t_1, t_2) = \sigma^2 \exp \left\{ - h_{\text{KS}}(L_1, L_2; l_1, l_2; t_1, t_2) \right\},
\label{eqn:kuusela_covariance}
\end{equation}
where 
\begin{equation*}
h_{\text{KS}}(L_1, L_2; l_1, l_2; t_1, t_2) = \sqrt{\left( \frac{L_1 - L_2}{\theta_{\text{lat}}} \right)^2 + \left( \frac{l_1 - l_2}{\theta_{\text{lon}}} \right)^2 + \left( \frac{t_1 - t_2}{\theta_t} \right)^2},
\end{equation*}
and $\theta_{\text{lat}},\theta_{\text{lon}},\theta_t>0$ are range parameters.

Most studies discretize the vertical domain into fixed pressure levels and interpolate each layer independently, which requires preprocessing irregular profile data and can introduce systematic error \citep{kuusela2018locally,hu2024spatiotemporal}. To avoid this, \cite{yarger2022functional} treated Argo profiles as functions of pressure and modeled the local mean as
\begin{eqnarray}
\mu_{L_0,l_0,t_0}(L,l,t,y,p)
&=&
\sum_{\tilde{y}=2007}^{2016}
\beta_{0,\tilde{y}}(p)\,\mathbbm{1}(y=\tilde{y})
+ (L-L_0)\,\beta_1(p)
+ (l-l_0)\,\beta_2(p)
\nonumber\\[4pt]
&&+ (L-L_0)^2\,\beta_3(p)
+ (l-l_0)^2\,\beta_4(p)
+ (L-L_0)(l-l_0)\,\beta_5(p)
\nonumber\\[4pt]
&&+ (t-t_0)\,\beta_6(p)
+ (t-t_0)^2\,\beta_7(p),
\label{eqn:yarger_mean}
\end{eqnarray}
where $\mathbbm{1}(\cdot)$ is an indicator function and $\beta_{0,\tilde{y}}(p)$ and $\beta_k(p)$ are smooth functions of pressure. This functional approach reduces prediction error and avoids gridding artifacts; the resulting residuals were analyzed via functional principal components.

\begin{figure}[bt!]
	\centering
	\includegraphics[scale=0.2]{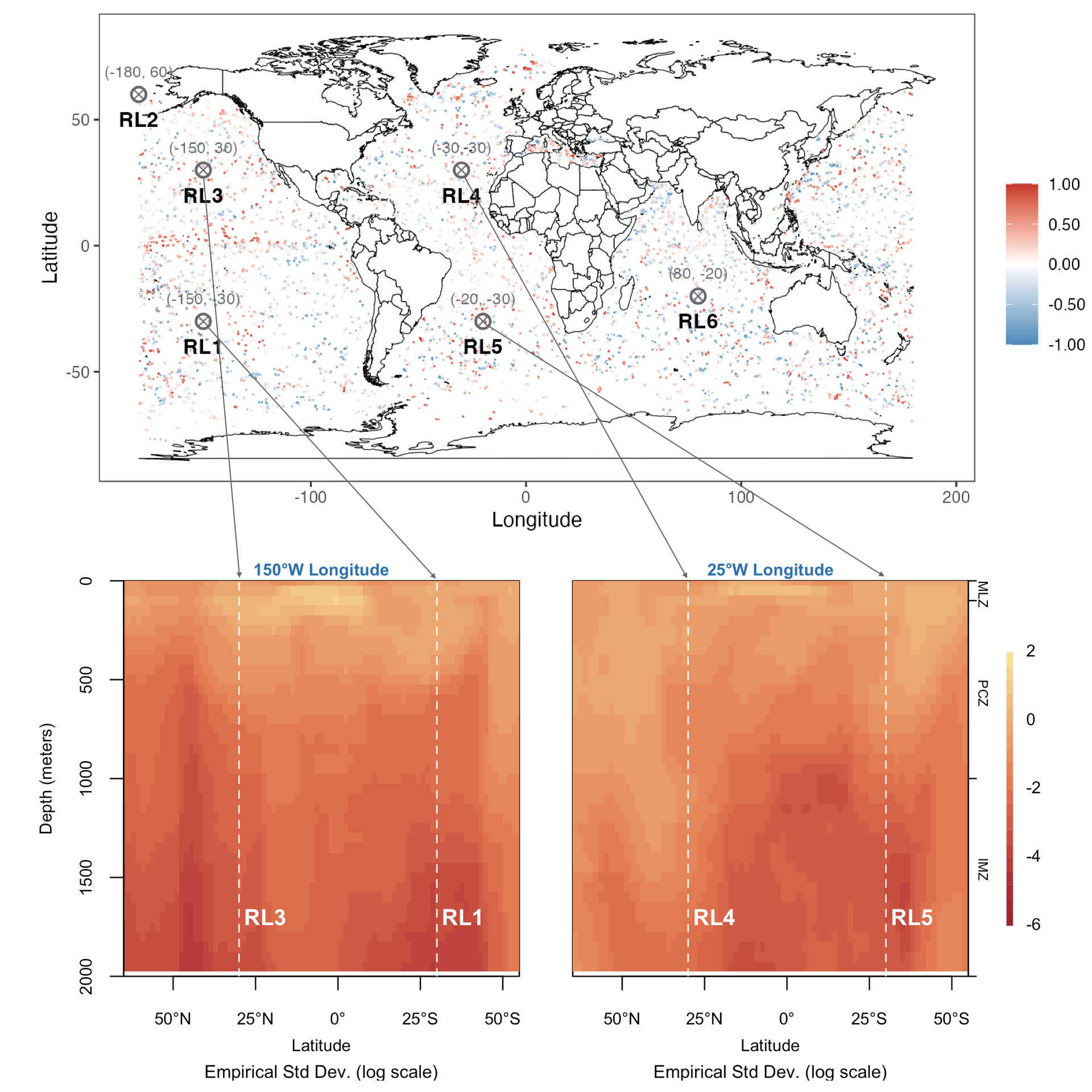}
    \caption{Temperature residuals from January to March in 2016 obtained by \cite{yarger2022functional}. (Top) Six reference locations are marked with their coordinates. \rev{Residuals are shown in their natural units ($^\circ C$) with the color scale truncated to $\pm 1$$^\circ C$  for visual clarity; a small number of more extreme values are omitted. No normalization or standardization was applied.} (Bottom) Empirical standard deviations (in log-scale) at every 50-meter depth interval along two selected longitude transects: 150\textdegree W and 25\textdegree W. These transects were chosen because they intersect the most number of reference locations while avoiding land crossings, allowing for a continuous vertical slice of the ocean to be examined. The latitudes of the closest reference locations along each transect are delineated with dashed lines and labeled accordingly.}
	\label{fig:yarger_residuals_temp}
\end{figure}

\begin{figure}[bht!]
	\centering
	\includegraphics[scale=0.2]{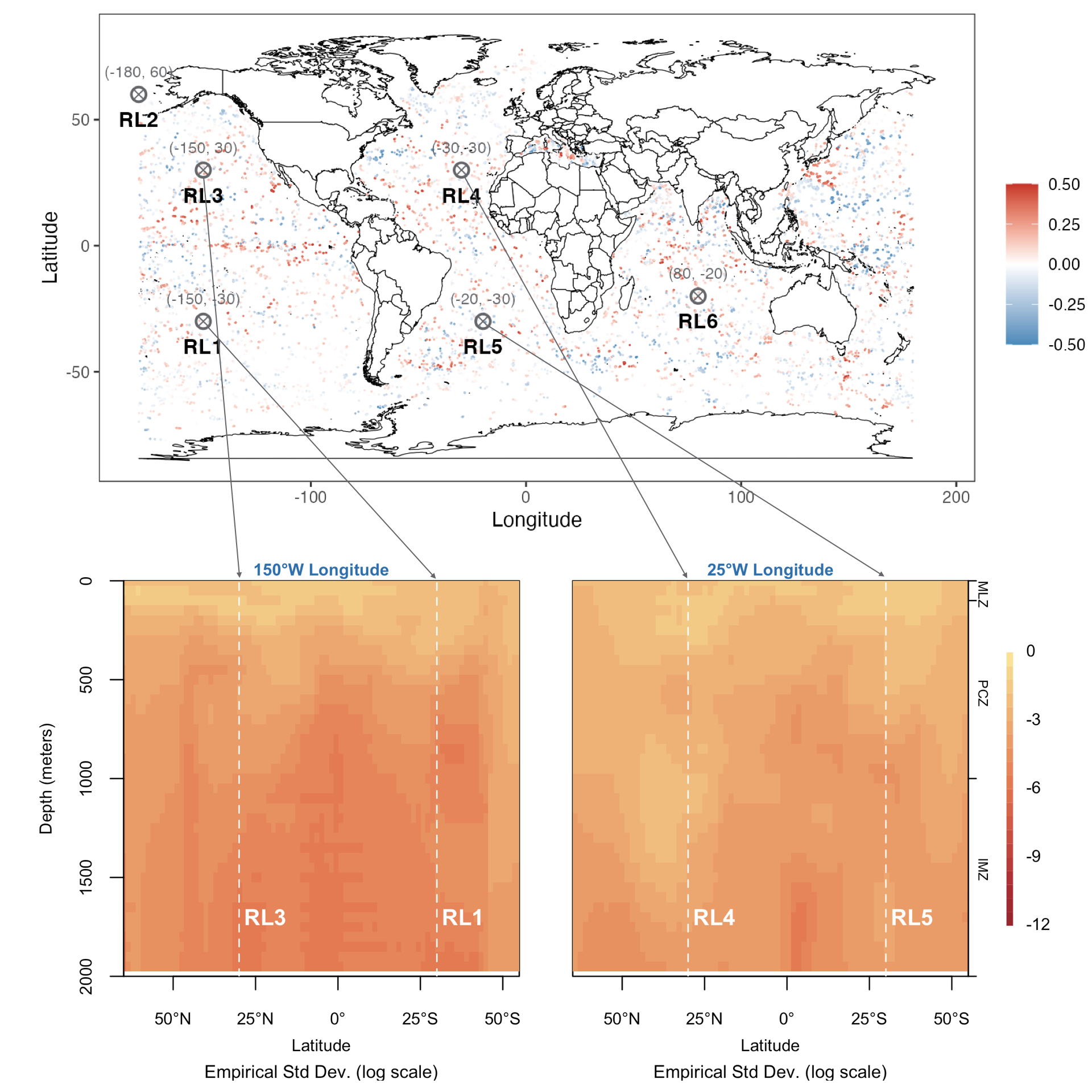}
    \caption{Same as Figure~\ref{fig:yarger_residuals_temp} but for salinity residuals. \rev{Residuals are in PSU, with the color scale truncated to $\pm 0.5$ PSU for legibility. No normalization or standardization was applied.}}
    \label{fig:yarger_residuals_psal}
\end{figure}

In our work, we focus on modelling the spatial dependence between the temperature and salinity residuals produced by this functional mean model. This enables a clearer characterization of their vertical covariability and supports improved prediction at unsampled locations. Figures~\ref{fig:yarger_residuals_temp} and \ref{fig:yarger_residuals_psal} show temperature and salinity residuals at the surface for January–March 2016. \rev{Following \citet{yarger2022functional}, we focus on this three-month window to limit temporal correlation, assuming stationarity over this short period.} For subsequent analysis, we designate six {\it reference locations} (RL), namely RL1: $(30^{\circ}S, 150^{\circ}W)$, RL2: $(60^{\circ}N, 180^{\circ}W)$, RL3: $(30^{\circ}N, 150^{\circ}W)$, RL4: $(30^{\circ}N, 30^{\circ}W)$, RL5: $(30^{\circ}S, 20^{\circ}W)$, and RL6: $(20^{\circ}S, 80^{\circ}E)$, for subsequent analyses. These locations were selected to represent major ocean gyres across the Pacific, Atlantic, and Indian Oceans. We also indicate three key vertical zones of ocean stratification: the mixed layer zone (MLZ, 0–100 m), the pycnocline zone (PCZ, 100–1{,}000 m), and the intermediate zone (IMZ, 1{,}000–2{,}000 m), also referred to as the deep layer \citep{chen2019mirror}. Since pressure in dbar is approximately equal to depth in meters, we use pressure as a vertical coordinate throughout.

Surface residuals show no strong large-scale spatial dependence patterns; values near the equator and along coasts often change abruptly, suggesting weaker dependence driven by localized dynamics or boundary effects. Over the open ocean, residuals vary more smoothly, indicating stronger spatial dependence. Apart from these features, no other prominent surface patterns are evident.

More complex patterns emerge below the surface. \rev{For each location $\mathbf{s}$, let $\tilde{\mathcal{D}}$ denote a surrounding neighborhood with cardinality $|\tilde{\mathcal{D}}|$.} Using 
\begin{equation*}
    \hat{\sigma}_i^2(\mathbf{s}) 
= \frac{\sum_{\tilde{\mathbf{s}}\in\tilde{\mathcal{D}}} \{Z_i(\tilde{\mathbf{s}})-\bar{\mu}_i(\mathbf{s})\}^2}{|\tilde{\mathcal{D}}|}
,\qquad
\hat{\rho}_{TS}(\mathbf{s})
= \frac{\sum_{\tilde{\mathbf{s}}\in\tilde{\mathcal{D}}}
\{T(\tilde{\mathbf{s}})-\bar{\mu}_T(\mathbf{s})\}
\{S(\tilde{\mathbf{s}})-\bar{\mu}_S(\mathbf{s})\}/|\tilde{\mathcal{D}}|}
{\hat{\sigma}_T(\mathbf{s})\,\hat{\sigma}_S(\mathbf{s})},
\end{equation*}
we compute empirical variances and colocated correlations for temperature ($T$) and salinity ($S$) along the $150^\circ\mathrm{W}$ and $25^\circ\mathrm{W}$ longitude lines on a $1^\circ \times 1^\circ \times 50$-m grid. The empirical mean is obtained using $\bar{\mu}_i(\mathbf{s}) = 
\frac{1}{|\tilde{\mathcal{D}}|}
\sum_{\tilde{\mathbf{s}}\in\tilde{\mathcal{D}}} Z_i(\tilde{\mathbf{s}}),
\;\rev{i\in\{T,S\}}$. These calculations require measurements of $Z_i$ at all $\tilde{\mathbf{s}}\in\tilde{\mathcal{D}}$. Here, $\tilde{\mathcal{D}}$ is taken as a vertical cylinder of radius 900 km and depth 50 m. Empirical variance estimates are shown in the bottom panels of Figures~\ref{fig:yarger_residuals_temp} and \ref{fig:yarger_residuals_psal}, and empirical colocated correlations appear in Figure~\ref{fig:empirical_colocated_correlation_vertical}.

The empirical variances reveal a clear depth dependence: variability is generally largest near the surface and decreases with depth. \rev{Local maxima commonly appear around 200–300 m, followed by a sharp decline below these depths} -- patterns consistent with prior studies \citep{mcphaden1991variability, chen2016vertical, chen2018climatology, al2020internal, xie2000interdecadal}. These structures align with MLZ, PCZ, and IMZ. In the MLZ (0–100 m), wind-driven mixing produces relatively uniform temperature and salinity, leading to low vertical variance \citep{helber2012temperature, maes2014seasonal}. The PCZ (100–1,000 m), coinciding with the thermocline and halocline, exhibits the highest variances, driven by mesoscale and submesoscale activity \citep{dong2014global, chen2016vertical, chen2018climatology}. Below this lies the IMZ (1,000–2,000 m), a more homogeneous and stable layer, reflected in persistently low variance.

Figure~\ref{fig:empirical_colocated_correlation_vertical} illustrates how temperature and salinity covary throughout the water column. Although temperature generally decreases and salinity increases with depth, their correlation is not simply monotonic. Near the surface, the variables are weakly correlated. Approaching the PCZ, their correlation becomes strongly positive, reaching a maximum between about 200 and 500 m. Below the PCZ, behavior diverges by longitude. Along $150^\circ\mathrm{W}$, correlations weaken with depth and eventually become strongly negative in the IMZ -- a pattern characteristic of the ``mirror layer'' described by \cite{chen2019mirror}. Along $25^{\circ}W$ longitude, however, strong positive correlations persist throughout the full 0–2{,}000 m range, consistent with the influence of the Atlantic meridional overturning circulation and density-compensating mechanisms that tightly couple temperature and salinity.

\subsection{Differential Operators Approach in 2D}

A class of 2D nonstationary spatial cross-covariance functions based on differential operators was introduced in \cite{jun2011non}, extending earlier univariate space–time work in \cite{jun2007approach}. In this framework, a multivariate spatial process is expressed as
\begin{equation} \label{eqn:jun_multivariate_process}
Z_i (L, l) = \sum_{k = 1}^K \left\{ a_{i, k}(L) \frac{\partial}{\partial L}  + b_{i, k}(L) \frac{\partial}{\partial l} \right\} X_k (L, l) + d_{i}(L) X_0 (L, l), \quad i = 1, \ldots, q,
\end{equation}
where $X_k (L, l)$, $k = 0 \ldots, K$, are independent zero-mean univariate spatial processes, and $a_{i, k}(L)$, $b_{i, k}(L)$, and $d_{i}(L)$ are non-random functions of the latitude. The resulting cross-covariance function, \text{cov}\{$Z_i (L_1, l_1), Z_j (L_2, l_2)\}$ may or may not have a closed form, depending on the choice of covariance functions for the latent processes $X_k$. Nonetheless, the formulation is flexible and accommodates negative covariances and spatially varying colocated correlations. 

A limitation of this representation is that all variables $Z_i$ depend on the same set of latent processes $X_k$. Greater flexibility can be obtained by replacing each $X_k$ with a multivariate latent field $\mathbf{X}_k(L,l)=\{X_{1,k}(L,l),\ldots,X_{q,k}(L,l)\}^{\top}$, such that the $Z_i$s depend on correlated but distinct components. This allows each $Z_i$ to have its own latent process while retaining cross-variable dependence through the latent multivariate fields. In the next section, we extend this differential operator framework to 3D and introduce nonstationarity along the vertical dimension, enabling construction of flexible bivariate cross-covariance models suited to Argo data.

\section{Methodology} \label{sec:methodology}

Empirical results show that both variances and colocated correlations of the residuals vary substantially with depth, and that vertical structure differs across horizontal locations. To accommodate this heterogeneity, we develop nonstationary cross-covariance models defined in 3D, capable of capturing the depth-dependent dependence patterns observed in the Argo residuals.

\subsection{Multivariate Nonstationary Spatial Models in 3D Domain}
\label{sec:model}

We treat the temperature and salinity residuals as two zero-mean 3D processes (latitude $\times$ longitude $\times$ depth) with spatial dependence encoded through a nonstationary cross-covariance function $C_{ij}(\mathbf{s}_1, \mathbf{s}_2) = \text{cov}\left\{ Z_i (\mathbf{s}_1), Z_j (\mathbf{s}_2)\right\}$, $i, j = 1, 2$, where $\mathbf{s}_1$ and $\mathbf{s}_2$ lie in 3D space. As usual, $C_{ij}$ must be positive definite to define a valid multivariate model.

To extend the 2D differential-operator construction of (\ref{eqn:jun_multivariate_process}) to 3D, we introduce a set of latent multivariate Gaussian processes.
For $k = 0,\ldots,K$, let
\begin{equation*}
    \mathbf{X}_k(L,l,p)
= \bigl(X_{1,k}(L,l,p),\ldots,X_{q,k}(L,l,p)\bigr)^\top
\end{equation*}
denote independent, zero-mean, $q$-variate stationary Gaussian processes with stationary cross-covariance
\[
C_{ij}^X(\mathbf{h}) 
= \operatorname{cov}\{X_{i,k}(\mathbf{s}), X_{j,k}(\mathbf{s}+\mathbf{h})\},
\]
common across $k$. These latent processes form the building blocks of the observed field $\mathbf{Z}$ through spatial differential operators. Incorporating an additional derivative in the vertical dimension, we obtain the following 3D extension:
\begin{prop} \label{theo:main_prop}
Let $\mathbf{X}_k (L, l, p)$, $k = 0, \ldots, K$, be the latent multivariate stationary processes defined above, each with cross-covariance $C_{ij}^X$. For $i = 1,\ldots,q$, let $a_{i,k}(L)$, $b_{i,k}(L)$, $c_{i,k}(L,p)$, and $d_i(L)$ denote deterministic coefficient functions that weight the differential operators. Define a multivariate process as follows:
\begin{equation} \label{eqn:jun_multivariate_process_proposed}
Z_i (L, l, p) = \sum_{k = 1}^K \left\{ a_{i, k}(L) \frac{\partial}{\partial L} + b_{i, k}(L) \frac{\partial}{\partial l} + c_{i, k}(L, p) \frac{\partial}{\partial p} \right\} X_{i,k} (L, l, p) + d_{i}(L) X_{i,0} (L, l, p).
\end{equation}
Furthermore, suppose the cross-covariance function $C_{ij}^X$ is chosen to be the parsimonious Mat\'{e}rn cross-covariance function in (\ref{eqn:multivariate_matern}) with smoothness $\nu_{ij} + 1$. Then, the multivariate process $\mathbf{Z} = (Z_1, Z_2, \ldots, Z_q)^{\top}$ has a nonstationary cross-covariance function:
\begin{multline} \label{eqn:proposed_cross_cov_explicit}
C_{ij}(L_1, L_2, l_1 - l_2, p_1, p_2) = K_{ij}^{1} \mathcal{M}_{\nu_{ij} - 1} \{ {h(L_1, L_2, l_1 - l_2, p_1 - p_2)}^{1/2} \} \\
+ K_{ij}^{2} \mathcal{M}_{\nu_{ij}} \{ {h(L_1, L_2, l_1 - l_2, p_1 - p_2)}^{1/2} \},
\end{multline}
where 
\begin{multline*}
K_{ij}^{1}  = \frac{\alpha_{ij}}{4} \sum_{k = 1}^K \{ a_{i,k} (L_1) a_{j, k} (L_2) h_1 h_2 - b_{i, k} (L_1) b_{j, k} (L_2) h_3^2 - c_{i, k} (L_1, p_1) c_{j, k} (L_2, p_2) h_4^2 \\- a_{i, k} (L_1) b_{j, k} (L_2) h_1 h_3 
+ a_{j, k} (L_2) b_{i, k} (L_1) h_2 h_3 - a_{i, k} (L_1) c_{j, k} (L_2, p_2) h_1 h_4 \\+ a_{j, k} (L_2) c_{i, k} (L_1, p_1) h_2 h_4  
 - b_{i, k} (L_1) c_{j, k} (L_2, p_2) h_3 h_4 - b_{j, k} (L_2) c_{i, k} (L_1, p_1) h_3 h_4 \}\\ + \alpha_{ij} h d_i (L_1) d_j (L_2), \\
K_{ij}^{2}  = - \frac{\alpha_{ij}}{2}  \sum_{k = 1}^K \{ a_{i, k} (L_1) a_{j, k} (L_2) h_{12} - b_{i, k} (L_1) b_{j, k} (L_2) h_{33} - c_{i, k} (L_1, p_1) c_{j, k} (L_2, p_2) h_{44} \\- a_{i, k} (L_1) b_{j, k} (L_2) h_{13} 
 + a_{j, k} (L_2) b_{i, k} (L_1) h_{23} \} + 2 \alpha_{ij} \nu_{ij} d_i (L_1) d_j (L_2),
\end{multline*}
and $\alpha_{ij} = \displaystyle\frac{\rho_{ij} \sigma_{i} \sigma_{j}}{2^{\nu_{ij}} \Gamma(\nu_{ij} + 1)}$. 
\end{prop} 

\noindent The proof of the proposition is rather straightforward. Refer to the Appendix in \cite{jun2007approach} for the exact expressions of $h_r$ and $h_{rs}$, for $r, s = 1, \dots, 4$. 

Proposition~\ref{theo:main_prop} extends the original 2D model in \eqref{eqn:jun_multivariate_process} by incorporating depth explicitly, enabling covariance structures that vary with vertical stratification—a dominant feature of ocean dynamics. It also introduces substantial structural flexibility: unlike the original formulation where all variables share the same latent processes (yielding colocated correlations of one), the proposed model allows each $Z_i$ to be driven by its own latent processes $X_{i,k}$. Dependence arises through cross-covariance among the latent fields, so colocated correlations $\rho_{ij}$ can be less than one, allowing more realistic relationships between variables such as temperature and salinity. Moreover, the differential operator structure accommodates spatial nonstationarity in both latitude and depth. The vertical derivative term $\partial/\partial p$ with depth-varying coefficient $c_{i,k}(L,p)$ enables the covariance structure to evolve with depth, capturing stratification, thermocline behavior, and other depth-dependent oceanographic processes. In practice, we focus on vertical nonstationarity within localized regions: the model is fit separately at each reference location, with $a_{i,k}$ and $b_{i,k}$ treated as local constants and $c_{i,k}(p)$ allowed to vary with depth. This targets model flexibility to the vertical dimension, where the strongest and most systematic heterogeneity occurs.

\subsection{Modelling vertical coefficients for the differential operator}
\label{sec:basis_choice}

We now present our strategies for modelling \( c_{i,k}(p) \) in a parsimonious and interpretable way. Since no constraints on \( c_{i,k}(p) \) are required for positive definiteness, these functions can be freely parameterized to introduce vertical nonstationarity.  
Earlier 2D work \citep{jun2008nonstationary, jun2011non} incorporated nonstationarity in latitude via spherical harmonics; here we adapt this idea to depth by representing \( c_{i,k}(p) \) through basis expansions. \rev{Although the overall model is 3D, the vertical direction is treated distinctly: \( c_{i,k}(p) \) governs the strength of the vertical derivative and varies smoothly with depth, enabling anisotropy and nonstationarity along the vertical axis.}

To represent \( c_{i,k}(p) \), we use a basis expansion 
\begin{align}
c_{i,k}(p) = \sum_{m = 0}^{M_i} \eta_{i,m}^{(k)} B_{i,m}(p), \label{c}
\end{align}
where \( \eta_{i,m}^{(k)} \) are coefficients and \( B_{i,m}(p) \) are basis functions defined over the normalized depth domain \( p \in [0, 1] \). Following the construction in \cite{jun2008nonstationary} and \cite{jun2011non}, we consider Legendre polynomials as a natural choice for \( B_{i,m} \). If \( P_m(x) \) denotes the \( m \)th Legendre polynomial defined on \( x \in [-1, 1] \), we define \( B_{i,m}(p) = P_m(2p - 1) \) after rescaling the depth variable \( p \) to the unit interval. This orthogonal basis provides smooth global representations well suited for broad vertical structures.
As an alternative, B-spline basis functions \( \mathcal{B}_m(p) \) may be used. B-splines offer localized control by joining piecewise polynomials at knots \citep{kunoth2018splines}. Let \( \mathcal{B}_m(p) \) denote the \( m \)th B-spline basis function of a chosen order. Then the basis expansion is written as \( B_{i,m}(p) = \mathcal{B}_m(p) \). The placement and spacing of knots can be tailored to the variable of interest or the structure of the data, with closely spaced knots enabling finer resolution to capture sharp vertical transitions. However, using a large number of knots increases the number of parameters and can complicate estimation due to potential convergence issues.

Both Legendre polynomials and B-splines are capable of producing flexible nonstationary covariance structures along the vertical depth axis. For our application, we found that Legendre polynomial bases provided a better overall fit when comparing models with an equal number of parameters (i.e., same polynomial order versus number of spline knots). This basis will therefore be used in the subsequent modelling.

\begin{figure}[tb!]
	\centering
		\includegraphics[scale=0.4]{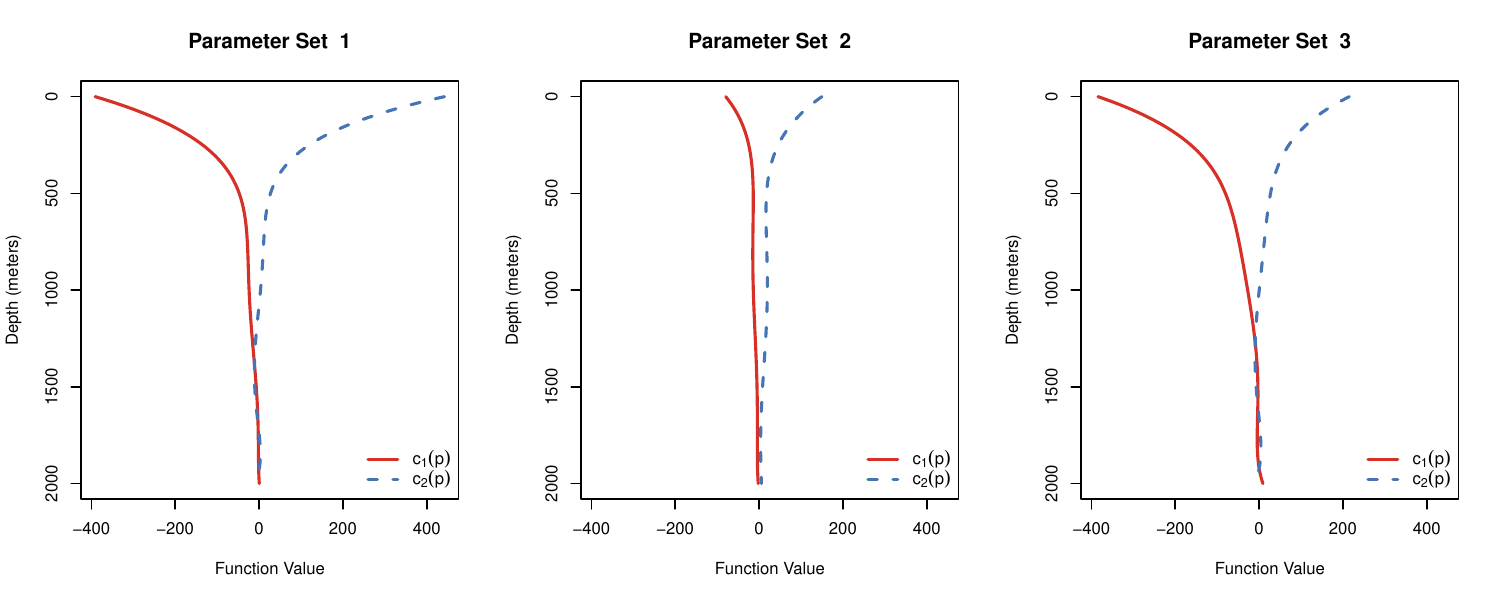}	
        
               \hspace*{1.cm}\includegraphics[scale=0.35]{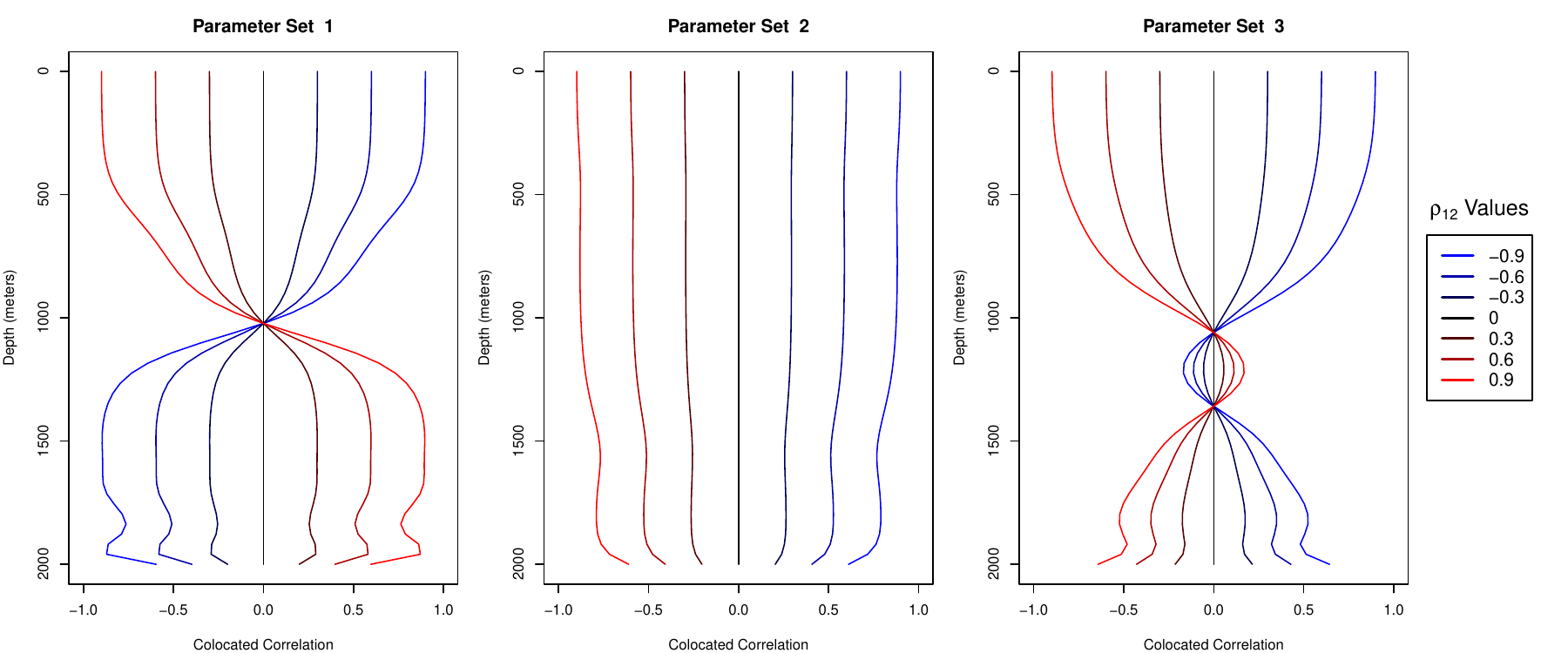}	
    \caption{Vertical shapes of $c_i(p)$ for three sets of coefficients (top row) and resulting colocated correlation curves (bottom row). Colocated correlations are shown for \( \rho_{12} = -0.9, \ldots, 0.9 \).}
    \label{fig:c1_c2_vs_depth}
\end{figure}

The functions \( c_{i,k}(p) \) directly shape vertical variability, influencing both marginal variances \( C_{ii}(L,L,0,p,p) \) and colocated cross-covariances \( C_{ij}(L,L,0,p,p) \). To illustrate the effect of these coefficients, consider the colocated correlation,
\[
\frac{C_{ij}(L, L, 0, p, p)}{\sqrt{C_{ii}(L, L, 0, p, p) \, C_{jj}(L, L, 0, p, p)}},
\]
in a bivariate setting (\( q=2 \)) with \( d_1=d_2=0 \). For parsimony, as in \cite{jun2007approach,jun2008nonstationary}, we drop the index \( k \) and refer to the coefficients simply as \( a_i \), \( b_i \), and \( c_i(p) \). The functions \( c_1(p) \) and \( c_2(p) \) used in our illustration are degree-5 Legendre polynomials estimated at three reference locations (RL1, RL4, RL5). The top panels of Figure~\ref{fig:c1_c2_vs_depth} display these estimated vertical coefficients, showing clear variation in structure across locations. In each panel, the red curve corresponds to \( c_1(p) \), and the blue dashed curve to \( c_2(p) \).
Using these estimated functions, the bottom panel of Figure~\ref{fig:c1_c2_vs_depth} shows the resulting colocated correlation curves as a function of depth, for a range of values of the colocated correlation parameter \( \rho_{12} \in (-1, 1) \). These curves illustrate how the shape of vertical dependence is modulated by \( c_1(p) \) and \( c_2(p) \), while \( \rho_{12} \) controls the amplitude of the correlation.

\subsection{Likelihood Approximation}

For the estimation of parameters, we use maximum likelihood estimation (MLE) to jointly estimate all parameters.
Despite employing a local stationary framework, the size of the Argo residuals dataset renders the matrix operations needed in the likelihood computation infeasible. Specifically, the Gaussian likelihood requires the inversion of the covariance matrix that has a $\mathcal{O}(N^2q^2)$ memory complexity and a $\mathcal{O}(N^3q^3)$ computational complexity, where $N$ is the number of spatial locations.

Many methods have been proposed in the literature to approximate Gaussian likelihoods; see \cite{heaton2019case} and \cite{liu2020gaussian} for a comprehensive review of such approaches for spatial data. Among these, we adopt the Vecchia approximation \citep{vecchia_estimation_1988}.
The Vecchia approximation is typically applied to univariate Gaussian processes (i.e., $q = 1$) and is based on truncating the conditioning in the joint density decomposition:
\begin{align}
    f(\bx) &= \prod_{w = 1}^{N} f(x_{w} | \bx_{1:w-1})
    \approx \prod_{w = 1}^{N} f(x_{w} | \bx_{\tau(w)}), \label{eqn:univar_vecc}
\end{align}
where $\bx$ is a random vector of length $N$, $f(\cdot)$ is a generic density function, and $\tau(w)$ is the conditioning set of $x_{w}$, commonly chosen as the $u$ nearest neighbors of $x_{w}$ among $\{x_{1}, \ldots, x_{w - 1}\}$ \citep{katzfuss2021general}. The parameter $u$ serves as a tuning parameter that balances the computation cost with the bias introduced by the approximation and is usually set such that $u \ll N$ (e.g., $u = 30$ for $N = 5{,}000$). Finding nearest neighbors requires a distance function defined over the domain of the response variables, for which we use $\sqrt{h}$ with $h$ defined in \eqref{eqn:distance_formula} and computed at $a_{h} = a_{v} = 1$. The conditioning sets $\{\tau(w)\}$ may vary with $a_{h}$ and $a_{v}$ but their impact on parameter estimation is usually insignificant and any choice of $\{\tau(w)\}$ leads to unbiased estimators \citep{cao2022scalable}. In our implementation, we fix $\{\tau(w)\}$ during optimization to enable automatic differentiation and improve computation efficiency but note that \cite{katzfuss2021general} proposes to recalculate the conditioning sets $\{\tau(w)\}$ after a number of iterations during optimization. For multivariate Gaussian processes, we substitute the random variable $x_{w}$ with a random vector of length $q$, for example, $[z_{1}({\bf{s}}_{w}), z_{2}({\bf{s}}_{w}), \ldots, z_{q}({\bf{s}}_{w})]^{\top}$, representing the $q$ variables observed at the $w$-th spatial location, under which the complexity for likelihood estimation becomes $\mathcal{O}(Nu^3q^3)$.

\section{Application} \label{sec:application}

We now fit the proposed covariance functions to jointly model Argo temperature and salinity residuals and compare their performance with the classic parsimonious Mat\'ern model. The results demonstrate the ability of our approach to capture depth-dependent structure and nonstationary cross-correlation between the two variables.

\subsection{Data}
\label{sec:float_selection}

Due to the sheer size of the full Argo dataset and the horizontal nonstationarity of ocean variables, modelling is performed in one local region—i.e., one reference location—at a time. We analyze six RLs selected to represent major ocean circulation regimes, as introduced in Section~\ref{sec:background}. These locations (marked with circled X symbols in Figures~\ref{fig:yarger_residuals_temp} and~\ref{fig:yarger_residuals_psal}) span high, midlatitude, and subtropical regions and sample key longitudinal bands to capture the thermohaline diversity of the global ocean. Each RL lies within a distinct circulation system: RL1 and RL2 are situated within the North Pacific; RL3 represents the South Pacific; RL4 lies in the Gulf Stream–North Atlantic region; RL5 captures the Brazil Current; and RL6 samples the East Australian Current. This distribution facilitates comparison across different oceanographic environments and provides a basis for evaluating how well the proposed 3D model adapts to regions with different temperature–salinity variability and stratification characteristics.

\begin{table}[htb!]
\caption{Counts of available floats and observations per variable, for each reference location. 
}
\centering
\scalebox{1}{
\begin{tabular}{ c|cccc } 
\hline
\multirow{2}{*}{RL}& \multicolumn{2}{c}{Training}&\multicolumn{2}{c}{Testing}\\
\cline{2-5}
&\# floats&\# obs&\# floats&\# obs \\
\hline
1 & 29 & 14,913 & 7 & 5,051 \\ 
2 & 24 & 15,475 & 6 & 3,946 \\ 
3 & 79 & 25,688 & 20 & 7,001  \\ 
4 & 66 & 19,747 & 16 & 3,938 \\ 
5 & 59 & 24,553  & 15 & 7,570 \\ 
6 & 64 & 28,179 & 16 & 7,067 \\ 
\hline
\end{tabular}
}
\label{table:2}
\end{table}

\rev{We use the 2016 residuals obtained from the functional mean model of \cite{yarger2022functional}, located within a vertical cylinder of radius 900~km centered at each reference location.} Table~\ref{table:2} summarizes the number of floats and total measurements available per region. Because floats differ in the depth levels they sample, some profiles span the full 0–2{,}000 m range while others terminate around 1{,}200 m. For each RL, we randomly split the floats into training and testing sets using an 80–20 ratio.

Before modelling temperature and salinity data as a bivariate Gaussian random field, we assessed the Gaussianity of the residuals at all six RLs. Histograms and Q–Q plots (see Supplementary Material) indicate that, in most ocean sub-regions, the data do not show significant deviations from the Gaussian behavior. Prior studies have also modeled temperature and salinity as Gaussian processes \citep{bohme2005objective, cabanes2016improvement}. In cases where non-Gaussian behavior is observed, common practice in the literature involves preprocessing the data to remove outliers \citep{chaigneau2011vertical}. For the six RLs used in this study, we model the residuals without additional preprocessing, as we did not detect any strong violations of Gaussianity.

\subsection{Covariance models} 
\label{sec:cov_model}

For bivariate modelling, we compare three covariance models; definitions and parameter counts are summarized in Table~\ref{tab:model_properties_summary}.

A preliminary univariate analysis showed that fitted Mat\'ern smoothness parameters for both variables were consistently close to~1 across all six RLs. \rev{Based on this empirical finding, we fix the smoothness so that all models yield an effective smoothness of $\nu_{\text{eff}} = 1$. For M1 (BiMat\'ern), this means setting $\nu = 1$ directly. For M2 and M3 (DiffOp models), we set $\nu = 2$ in the latent processes; because the differential operator construction involves first-order derivatives, the effective smoothness of the resulting processes $Z_i$ is $\nu - 1 = 1$. Moreover, we set $d_i(L)=0$ to keep the model parsimonious, so each $Z_i$ is constructed solely from first–derivative terms.} To incorporate vertical nonstationarity, the coefficient \( c_i(L,p) \) is taken to depend on pressure only, denoted \( c_i(p) \), and represented using the basis expansion in \eqref{c} with Legendre polynomials. We experimented with various polynomial orders and observed negligible improvements in model fit beyond order 5. Accordingly, we used degree-5 Legendre polynomials for both variables (i.e., \( M_1 = M_2 = 5 \)) throughout our analysis. \rev{The nugget effect was assumed in previous works, for example, \citet{kuusela2018locally}, and provides the benefits of accommodating measurement errors as well as numerical singularities. In this paper, we include a `relative' nugget effect parameter $\epsilon^2$ for all the models that amounts to a nugget effect equal to the diagonal entries of the cross-covariance matrix multiplied by $\epsilon^2$, considering the change of variation across depth.} As shown in Table~\ref{tab:model_properties_summary}, this specification yields 21 parameters for M2 (DiffOp--Ind) and 22 parameters for M3 (DiffOp--Bi).

\begin{table}[htb!]
{
\caption{Summary of covariance models, parameters, and properties.}
\centering
\scalebox{0.92}{
\begin{tabular*}{\textwidth}{@{\extracolsep{\fill}} l p{12cm}}
\toprule
Model & Specification \\
\midrule

\multirow{4}{*}{\textbf{M1 (BiMatérn)}}
& \textbf{Covariance Function:} \\[0.5ex]
& $\displaystyle 
C_{ij}^{\text{M1}}(\mathbf{s}_1,\mathbf{s}_2)
= \alpha_{ij}\,\mathcal{M}_{\nu_{ij}}\!\big(\sqrt{h(\mathbf{s}_1,\mathbf{s}_2)}\big),
$ \\[1ex]
& $\displaystyle 
\alpha_{ij}=\frac{\rho_{ij}\sigma_i\sigma_j}{2^{\nu_{ij}-1}\Gamma(\nu_{ij})};
$ \\[1ex]
& \textbf{Parameter Vector:} $\displaystyle 
\theta_{\text{M1}}=(\sigma_1^2,\sigma_2^2,a_{\text{h}},a_{\text{v}},\rho_{12})
$ \\[1.0ex]
& \textbf{Effective smoothness:} $\nu_{\text{eff}} = 1$ \\[0.5ex]
& \textbf{Vertical nonstationarity:} no \\[0.5ex]
& \textbf{Cross-covariance:} yes (stationary Matérn) \\[0.5ex]
& \textbf{\# parameters:} 6 (including nugget) \\
\midrule

\multirow{6}{*}{\textbf{M2 (DiffOp--Ind)}}
& \textbf{Covariance Function:} \\[0.5ex]
& $\displaystyle 
C_{12}^{\text{M2}}(\mathbf{s}_1,\mathbf{s}_2) = 0,
$ \\[1ex]
& $\displaystyle
C_{ii}^{\text{M2}}(\mathbf{s}_1,\mathbf{s}_2)
= K_{ii}^{1}\,\mathcal{M}_{\nu_{ii}-1}\!\big(\sqrt{h(\mathbf{s}_1,\mathbf{s}_2)}\big)
+ K_{ii}^{2}\,\mathcal{M}_{\nu_{ii}}\!\big(\sqrt{h(\mathbf{s}_1,\mathbf{s}_2)}\big),
$ \\[1ex]
& $K_{ii}^{1},K_{ii}^{2}$ from Prop.~\ref{theo:main_prop} with $K=1$, $d_i = 0$; \\[1ex]
& $\displaystyle 
c_i(p)=\sum_{m=0}^5 \eta_{i,m} B_m(p),\quad B_m(p)=P_m(2p-1);
$ \\[1ex]
& \textbf{Parameter Vector:} $\displaystyle 
\theta_{\text{M2}}=(\sigma_1^2,\sigma_2^2,a_{\text{h}},a_{\text{v}},
a_1,b_1,\eta_{1,0{:}5},a_2,b_2,\eta_{2,0{:}5})
$ \\[1ex]
& \textbf{Effective smoothness:} $\nu_{\text{eff}} = \nu_{ii} - 1 = 1$ \\[0.5ex]
& \textbf{Vertical nonstationarity:} yes \\[0.5ex]
& \textbf{Cross-covariance:} no ($C_{12}=0$) \\[0.5ex]
& \textbf{\# parameters:} 21 (including nugget) \\
\midrule

\multirow{6}{*}{\textbf{M3 (DiffOp--Bi)}}
& \textbf{Covariance Function:} \\[0.5ex]
& $\displaystyle 
C_{ij}^{\text{M3}}(\mathbf{s}_1,\mathbf{s}_2)
= K_{ij}^{1}\,\mathcal{M}_{\nu_{ij}-1}\!\big(\sqrt{h(\mathbf{s}_1,\mathbf{s}_2)}\big)
+ K_{ij}^{2}\,\mathcal{M}_{\nu_{ij}}\!\big(\sqrt{h(\mathbf{s}_1,\mathbf{s}_2)}\big),
$ \\[1ex]
& $K_{ij}^{1},K_{ij}^{2}$ from Prop.~\ref{theo:main_prop}; \\[1ex]
& $\displaystyle 
c_i(p)=\sum_{m=0}^5 \eta_{i,m} B_m(p),\quad B_m(p)=P_m(2p-1);
$ \\[1ex]
& \textbf{Parameter Vector:} $\displaystyle
\theta_{\text{M3}}=(\sigma_1^2,\sigma_2^2,a_{\text{h}},a_{\text{v}},
a_1,b_1,\eta_{1,0{:}5},a_2,b_2,\eta_{2,0{:}5},\rho_{12})
$ \\[1ex]
& \textbf{Effective smoothness:} $\nu_{\text{eff}} = \nu_{ii} - 1 = 1$ \\[0.5ex]
& \textbf{Vertical nonstationarity:} yes \\[0.5ex]
& \textbf{Cross-covariance:} yes (nonstationary, depth-varying) \\[0.5ex]
& \textbf{\# parameters:} 22 (including nugget) \\

\bottomrule
\end{tabular*}
}
\label{tab:model_properties_summary}
}
\end{table}

\subsection{Fitted results}

We now examine the performance of the three candidate spatial covariance models, M1 (BiMat\'ern), M2 (DiffOp--Ind), and M3 (DiffOp--Bi), in modelling the residuals around the six RLs. As shown earlier in Figures~\ref{fig:yarger_residuals_temp}–\ref{fig:empirical_colocated_correlation_vertical}, each region exhibits a unique vertical dependence structure, and the ability of a model to reproduce these patterns varies accordingly. Model performance is assessed based on how well each model explains the observed residuals, using both log-likelihood values and the Akaike Information Criterion (AIC) as metrics.

\begin{table}[htb!]
\caption{Parameter estimates and performance summaries for M1 (BiMat\'ern), M2 (DiffOp-Ind), and M3 (DiffOp-Bi) using datasets collected at RL 1 through 6. For parameter estimates, values in parentheses indicate asymptotic standard errors. Higher log-likelihood and lower AIC values indicate better model fit. The best values per RL are shown in \textbf{bold}.}
\centering
\scalebox{0.8}{
\begin{tabular}{c| c |c c c c c |c c c}
   \toprule
Reference & \multirow{2}{*}{Model} & \multirow{2}{*}{$\hat{\rho}_{12}$} & horizontal & vertical & \multirow{2}{*}{$\hat{\sigma}_1$} & \multirow{2}{*}{$\hat{\sigma}_2$} & log-likelihood & AIC\\
Location & & & scale, $\hat{a}_h$ & scale, $\hat{a}_v$ & & & $(\times 10^3)$ & $(\times 10^3)$ &\\[0.5ex] \hline

\multirow{6}{*}{RL1} & \multirow{2}{*}{M1 (BiMat\'ern)} & 0.532 & 899 & 64.4 & 0.352 & 0.328 & \multirow{2}{*}{59.2} & \multirow{2}{*}{-118} \\
 & & (5.5e-03) & (84.4) & (0.29) & (1.87e-03) & (1.74e-03)\\
 & \multirow{2}{*}{M2 (DiffOp-Ind)} & -- & 357 & 14.2 & 4.19e-05 & 7.78e-05 & \multirow{2}{*}{78} & \multirow{2}{*}{-156} \\
 & & -- & (1.98) & (0.04) & (1.67e-07) & (3.96e-07)\\
 & \multirow{2}{*}{M3 (DiffOp-Bi)} & 0.895 & 291 & 75.3 & 2.22e-04 & 3.29e-04 & \multirow{2}{*}{\textbf{99.6}} & \multirow{2}{*}{\textbf{-199}} \\
 & & (4.7e-03) & (2.25) & (0.21) & (1.12e-06) & (1.73e-06)\\[0.5ex]\hline
\multirow{6}{*}{RL2} & \multirow{2}{*}{M1 (BiMat\'ern)} & 0.831 & 933 & 52 & 0.233 & 0.22 & \multirow{2}{*}{85.8} & \multirow{2}{*}{-172} \\
 & & (2.3e-03) & (62.2) & (0.24) & (9.49e-04) & (8.98e-04)\\
 & \multirow{2}{*}{M2 (DiffOp-Ind)} & -- & 503 & 75.5 & 9.01e-05 & 1.94e-04 & \multirow{2}{*}{119} & \multirow{2}{*}{-238} \\
 & & -- & (3.08) & (0.24) & (5.12e-07) & (1.10e-06)\\
 & \multirow{2}{*}{M3 (DiffOp-Bi)} & 0.953 & 447 & 75.9 & 2.28e-04 & 3.48e-04 & \multirow{2}{*}{\textbf{122}} & \multirow{2}{*}{\textbf{-244}} \\
 & & (1.6e-03) & (2.25) & (0.26) & (1.11e-06) & (1.69e-06)\\[0.5ex]\hline
\multirow{6}{*}{RL3} & \multirow{2}{*}{M1 (BiMat\'ern)} & 0.765 & 3005 & 54.3 & 0.366 & 0.423 & \multirow{2}{*}{98} & \multirow{2}{*}{-196} \\
 & & (2.3e-03) & (551) & (0.19) & (1.27e-03) & (1.46e-03)\\
 & \multirow{2}{*}{M2 (DiffOp-Ind)} & -- & 1041 & 81.6 & 6.26e-05 & 1.05e-04 & \multirow{2}{*}{131} & \multirow{2}{*}{-261} \\
 & & -- & (6.95) & (0.17) & (2.83e-07) & (4.72e-07)\\
 & \multirow{2}{*}{M3 (DiffOp-Bi)} & 0.786 & 1008 & 84.6 & 7.58e-05 & 1.05e-04 & \multirow{2}{*}{\textbf{133}} & \multirow{2}{*}{\textbf{-266}} \\
 & & (5.1e-03) & (6.13) & (0.18) & (3.17e-07) & (4.38e-07)\\[0.5ex]\hline
\multirow{6}{*}{RL4} & \multirow{2}{*}{M1 (BiMat\'ern)} & 0.912 & 949 & 43.1 & 0.447 & 0.264 & \multirow{2}{*}{92.6} & \multirow{2}{*}{-185} \\
 & & (1.1e-03) & (67.6) & (0.18) & (1.29e-03) & (7.6e-04)\\
 & \multirow{2}{*}{M2 (DiffOp-Ind)} & -- & 1010 & 89.9 & 2.29e-04 & 1.58e-04 & \multirow{2}{*}{86.6} & \multirow{2}{*}{-173} \\
 & & -- & (10.5) & (0.19) & (1.16e-06) & (8.03e-07)\\
 & \multirow{2}{*}{M3 (DiffOp-Bi)} & 0.993 & 352 & 62.8 & 1.16e-03 & 3.54e-04 & \multirow{2}{*}{\textbf{97}} & \multirow{2}{*}{\textbf{-194}} \\
 & & (2e-04) & (2.81) & (0.12) & (3.54e-06) & (1.08e-06)\\[0.5ex]\hline
\multirow{6}{*}{RL5} & \multirow{2}{*}{M1 (BiMat\'ern)} & 0.468 & 332 & 42.5 & 0.571 & 0.251 & \multirow{2}{*}{106} & \multirow{2}{*}{-211} \\
 & & (5e-03) & (8.62) & (0.15) & (2.43e-03) & (1.07e-03)\\
 & \multirow{2}{*}{M2 (DiffOp-Ind)} & -- & 558 & 75.4 & 1.43e-04 & 2.57e-04 & \multirow{2}{*}{152} & \multirow{2}{*}{-304} \\
 & & -- & (3.24) & (0.16) & (6.48e-07) & (1.16e-06)\\
 & \multirow{2}{*}{M3 (DiffOp-Bi)} & 0.917 & 531 & 79.9 & 8.65e-05 & 3.21e-04 & \multirow{2}{*}{\textbf{155}} & \multirow{2}{*}{\textbf{-310}} \\
 & & (2.8e-03) & (2.78) & (0.17) & (3.58e-07) & (1.33e-06)\\[0.5ex]\hline
\multirow{6}{*}{RL6} & \multirow{2}{*}{M1 (BiMat\'ern)} & -0.156 & 1961 & 64.9 & 0.489 & 0.534 & \multirow{2}{*}{82} & \multirow{2}{*}{-164} \\
 & & (6e-03) & (137) & (0.21) & (2.05e-03) & (2.24e-03)\\
 & \multirow{2}{*}{M2 (DiffOp-Ind)} & -- & 613 & 38.8 & 1.54e-04 & 1.17e-04 & \multirow{2}{*}{139} & \multirow{2}{*}{-278} \\
 & & -- & (4.89) & (0.07) & (6.42e-07) & (4.85e-07)\\
 & \multirow{2}{*}{M3 (DiffOp-Bi)} & 0.912 & 633 & 60 & 1.20e-04 & 1.61e-04 & \multirow{2}{*}{\textbf{144}} & \multirow{2}{*}{\textbf{-289}} \\
 & & (2.2e-03) & (3.84) & (0.12) & (4.57e-07) & (6.12e-07)\\
 \bottomrule
\end{tabular}
}
\label{tab:parameters_differential_operators_new}
\end{table}

The results are summarized in Table~\ref{tab:parameters_differential_operators_new}. 
Both the log-likelihood and AIC values demonstrate that the proposed bivariate model M3 (DiffOp-Bi) consistently outperforms M1 (BiMat\'ern) and M2 (DiffOp-Ind) across all six RLs. This improvement reflects the enhanced flexibility of M3 (DiffOp-Bi) in capturing both nonstationary marginal variance and nonzero cross-correlation between temperature and salinity, made possible by the differential operator parameterization. While M2 (DiffOp-Ind) incorporates vertical nonstationarity through flexible marginal variances, its assumption of zero cross-correlation generally limits its performance relative to M3 (DiffOp-Bi), although it still outperforms the simpler stationary model M1 (BiMat\'ern) in most cases. Notably, the estimated colocated cross-correlation \( \hat{\rho}_{12} \) values under M3 (DiffOp-Bi) are consistently larger than those from M1 (BiMat\'ern) (except at RL3), suggesting that M3 (DiffOp-Bi) better captures cross-correlation. In addition, the estimated horizontal and vertical scale parameters under M2 (DiffOp-Ind) and M3 (DiffOp-Bi) are broadly consistent with each other, yet they differ substantially from those of M1 (BiMat\'ern), which assumes stationarity and lacks flexibility in vertical structure.

\begin{figure}[t!]
	\centering
		\includegraphics[scale=0.65]{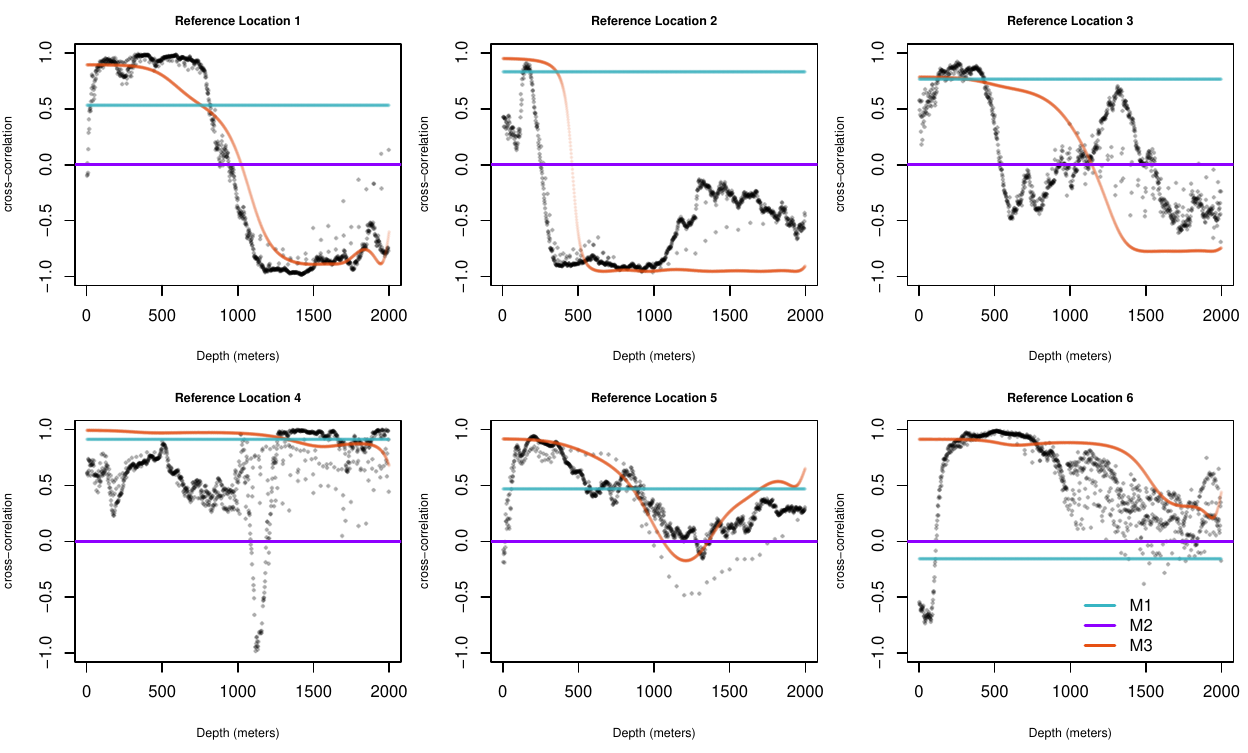} 
     \caption{Empirical (black) and model-based estimates of colocated correlations plotted as a function of depth for each RL. Empirical values are computed as sample cross-correlations of temperature and salinity from all training floats within non-overlapping vertical bins of 2-meter width. Model M1 (BiMat\'ern) produces constant correlations across depth; M2 (DiffOp-Ind) yields zero throughout; M3 (DiffOp-Bi) tracks the empirical patterns.}
     \label{fig:fitted_vs_empirical}
\end{figure}

Figure~\ref{fig:fitted_vs_empirical} displays the fitted colocated correlation curves produced by the three models, overlaid with empirical estimates from the training data (in black) as a function of depth. Empirical correlations are obtained by binning depth into 2-meter intervals and pooling all observations within each bin. Across all six RLs, the fitted curves from M3 (DiffOp-Bi) align closely with the empirical colocated correlations, successfully capturing the complex vertical dependence structure from the surface down to 2000 meters. While the fit is not perfect at RL2 and RL3, M3 (DiffOp-Bi) still demonstrates substantially better agreement with the data than the other models. Model M2 (DiffOp-Ind), which assumes independence between variables, fails to reflect the strong cross-correlation observed in all six regions. Similarly, the baseline stationary model M1 (BiMat\'ern) is unable to reproduce the depth-varying patterns, leading to a poor overall fit. These results highlight the inadequacy of stationary models like M1 (BiMat\'ern) for representing realistic vertical dependence and emphasize the utility of flexible, vertically nonstationary models such as M3 (DiffOp-Bi).

\subsection{Kriging results}

We next compare the three models by evaluating how well they predict the measurements of testing floats. As shown in Table~\ref{table:2}, 20\% of floats at each RL were randomly withheld for testing. For every testing float, we performed kriging for both temperature and salinity using observations from the five geographically closest training floats.

\begin{figure}[tb!]
  \centering
    \includegraphics[width=0.7\textwidth]{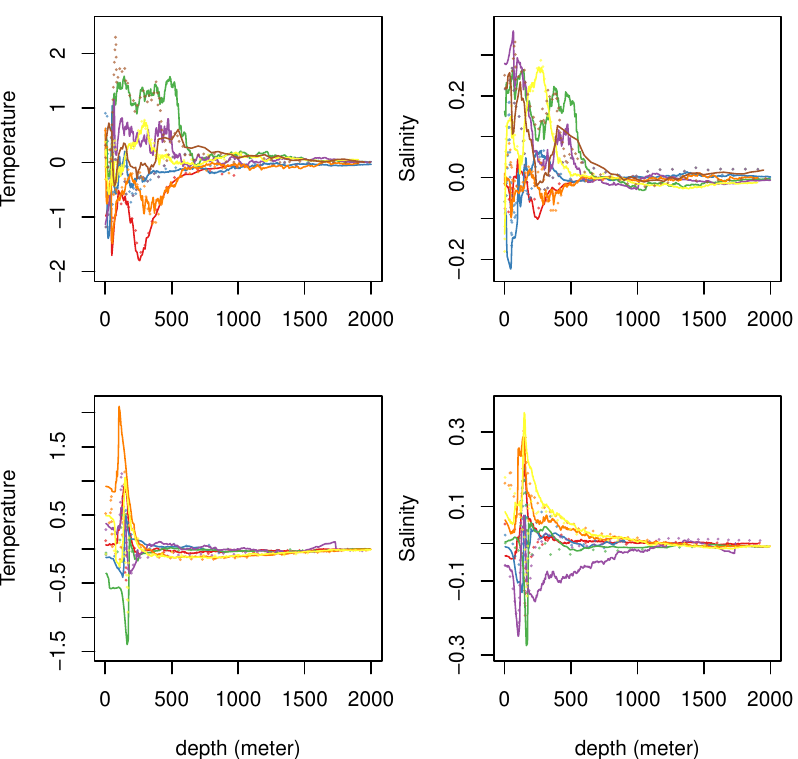}
    \caption{Comparison between residual profiles from testing floats and their (horizontally) nearest training floats at RL 1 (top row) and 2 (bottom row), for temperature residuals (left) and salinity residuals (right). For each testing float (solid curve), the corresponding nearest training float is displayed in the same color with a dotted line.}
  \label{fig:testing_float_nearest_neighbor}
\end{figure}

A major challenge in this prediction task is the strong variability in vertical structure across floats, even within the same local region. Figure~\ref{fig:testing_float_nearest_neighbor} illustrates this issue by comparing the temperature and salinity residual profiles of testing floats with those of their nearest training float (in horizontal distance) at RL1 (7 testing floats) and RL2 (6 testing floats). The remaining 4 RLs have more floats, and displaying and comparing profiles from each of the floats would be more difficult.  
Overall, the discrepancies between testing and training data are substantial for both regions and both variables. While in some cases the solid and dotted curves match closely, in many others they diverge significantly—especially within the upper 1,000 meters of the water column. These mismatches indicate either horizontal nonstationarity or weak horizontal correlation and highlight the difficulty of prediction, even when using covariance models explicitly designed to accommodate nonstationary structure. Notably, some of these discrepancies between nearby floats may stem from the fact that floats are displaced horizontally as they travel vertically. Argo float data report a single longitude–latitude coordinate per profile, although floats may drift laterally due to ocean turbulence and other dynamic factors while profiling \citep{wang2020eddy,merchel2024use}.
 
To assess the depth-dependent predictive performance of the models, we compute the Mean Absolute Error (MAE) of kriging predictions. MAE is calculated separately for each of the six RLs and for each of the two variables—temperature and salinity—across the three models.
Because each RL contains multiple testing floats, we summarize MAE values by binning depth into 100-meter intervals: [0,100), [100,200), $\ldots$, [1900,2000] meters. For each bin, we pool all prediction errors from testing profiles whose observation depths fall within that interval. This approach ensures adequate sample sizes within bins and yields stable MAE estimates across depth.
Specifically, let \( z^{\text{true}}_w \) and \( z^{\text{pred}}_w \) denote the true and predicted temperature or salinity residual at depth \( d_w \), and let \( \mathcal{D}_b = \{w : d_w \in \text{bin}_b \} \) denote the set of observations in depth bin \( b \). The MAE for that bin is:
\[
\text{MAE}_b = \frac{1}{|\mathcal{D}_b|} \sum_{w \in \mathcal{D}_b} \left| z^{\text{true}}_w - z^{\text{pred}}_w \right|.
\]

\begin{figure}[tb!]
  \centering
  \begin{minipage}[t]{0.45\textwidth}
    \centering
    \includegraphics[width=\textwidth]{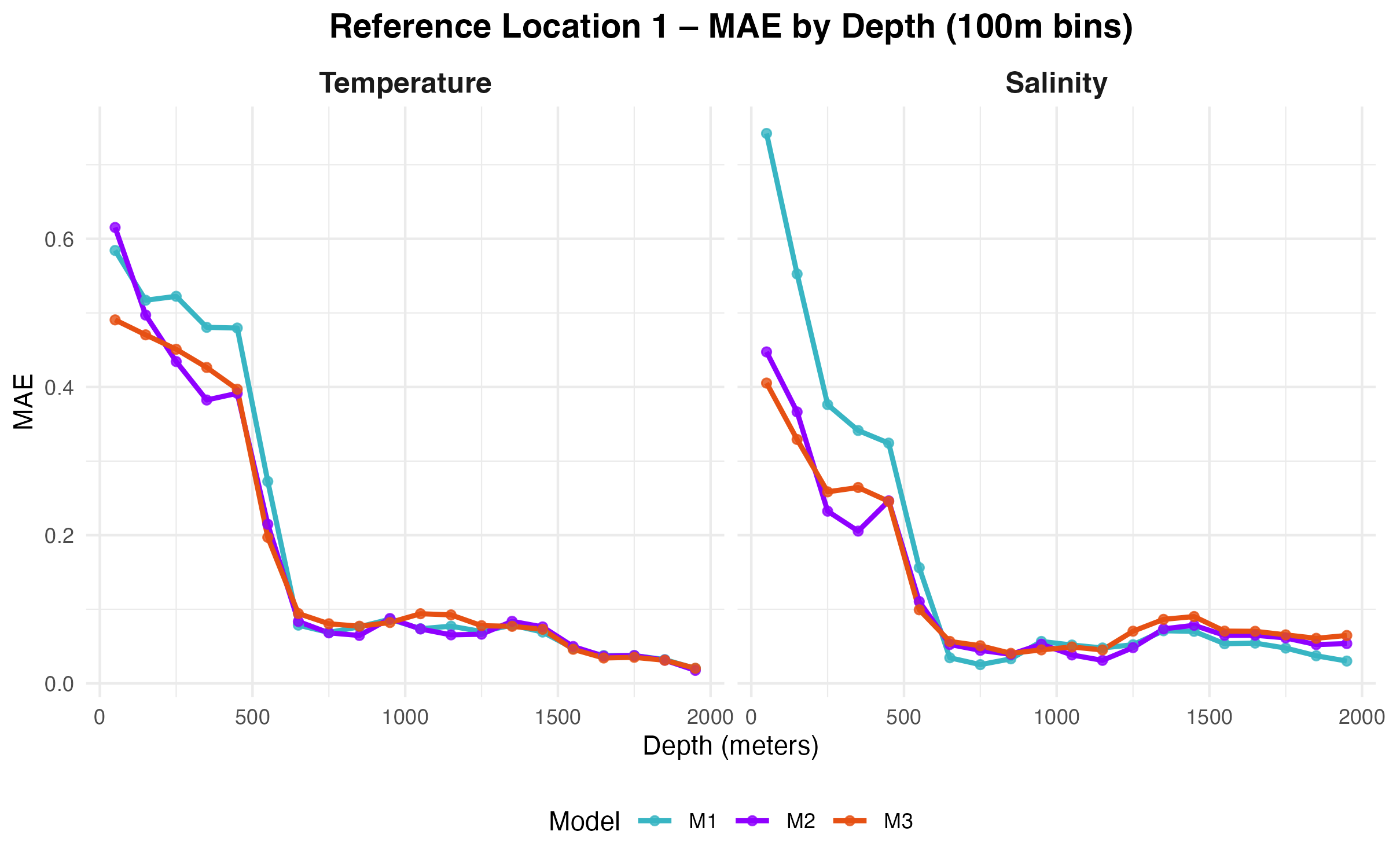}
  \end{minipage}%
  \hfill
  \begin{minipage}[t]{0.45\textwidth}
    \centering
    \includegraphics[width=\textwidth]{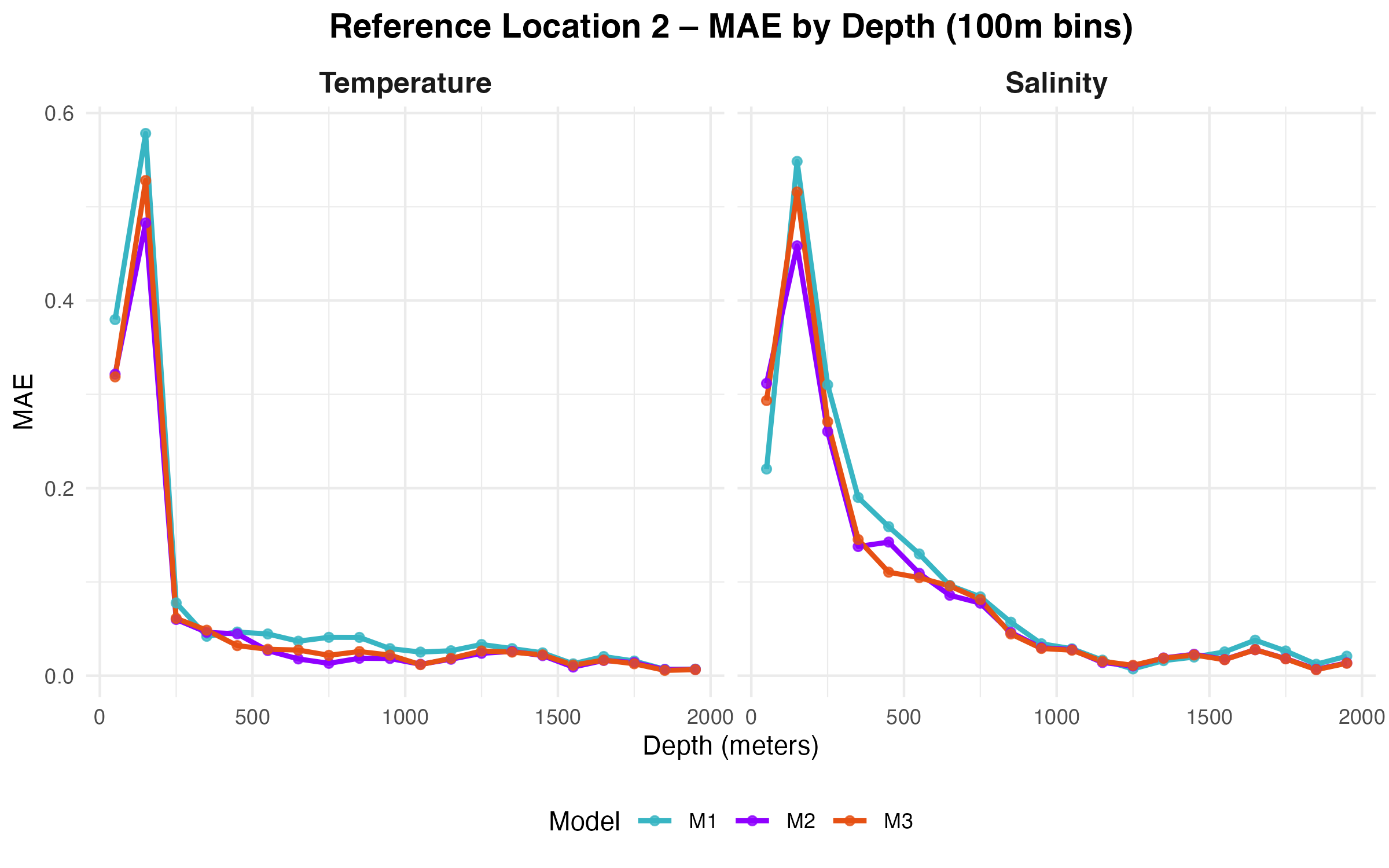}
  \end{minipage}

  \vspace{1em}
  \begin{minipage}[t]{0.45\textwidth}
    \centering
    \includegraphics[width=\textwidth]{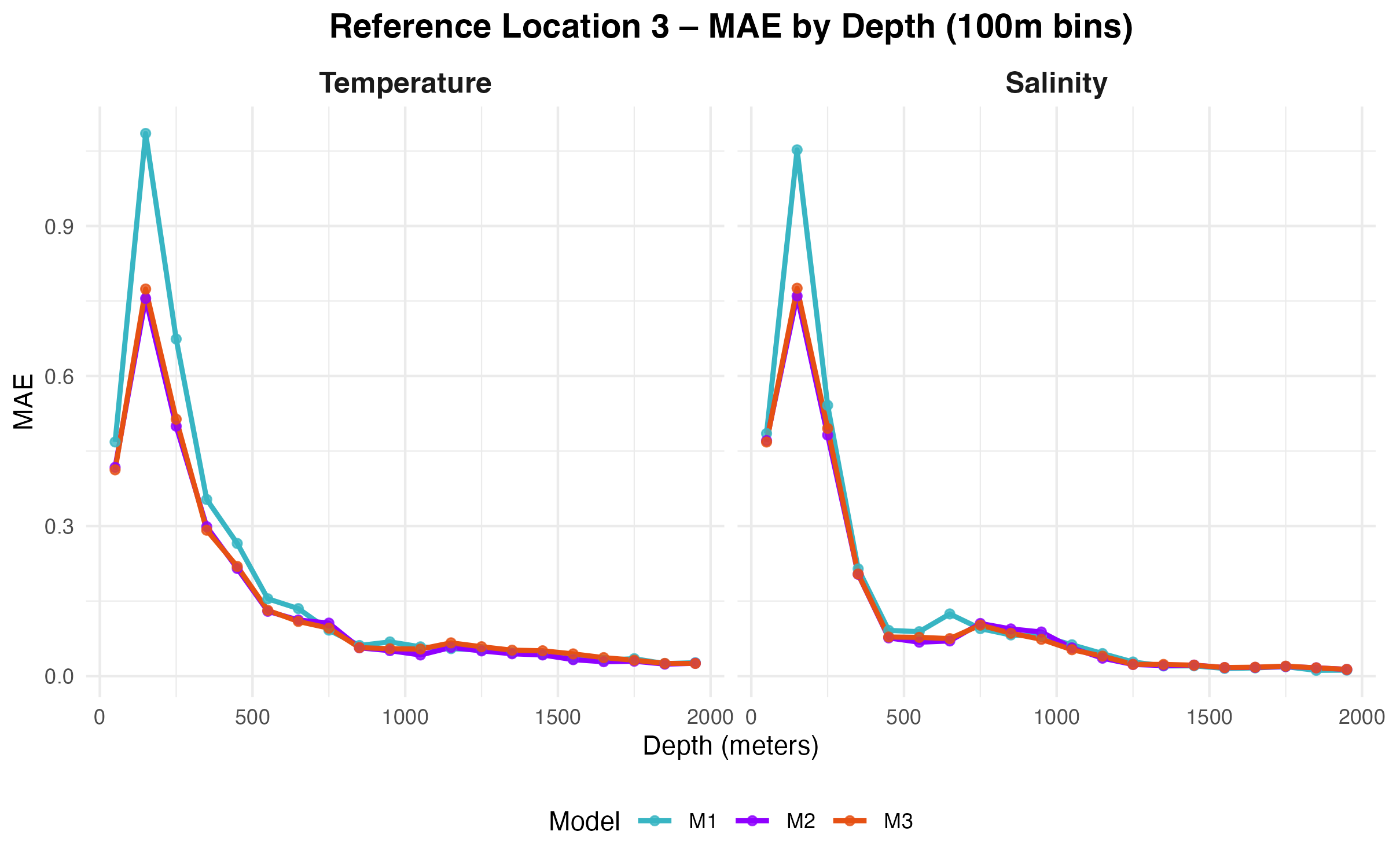}
  \end{minipage}%
  \hfill
  \begin{minipage}[t]{0.45\textwidth}
    \centering
    \includegraphics[width=\textwidth]{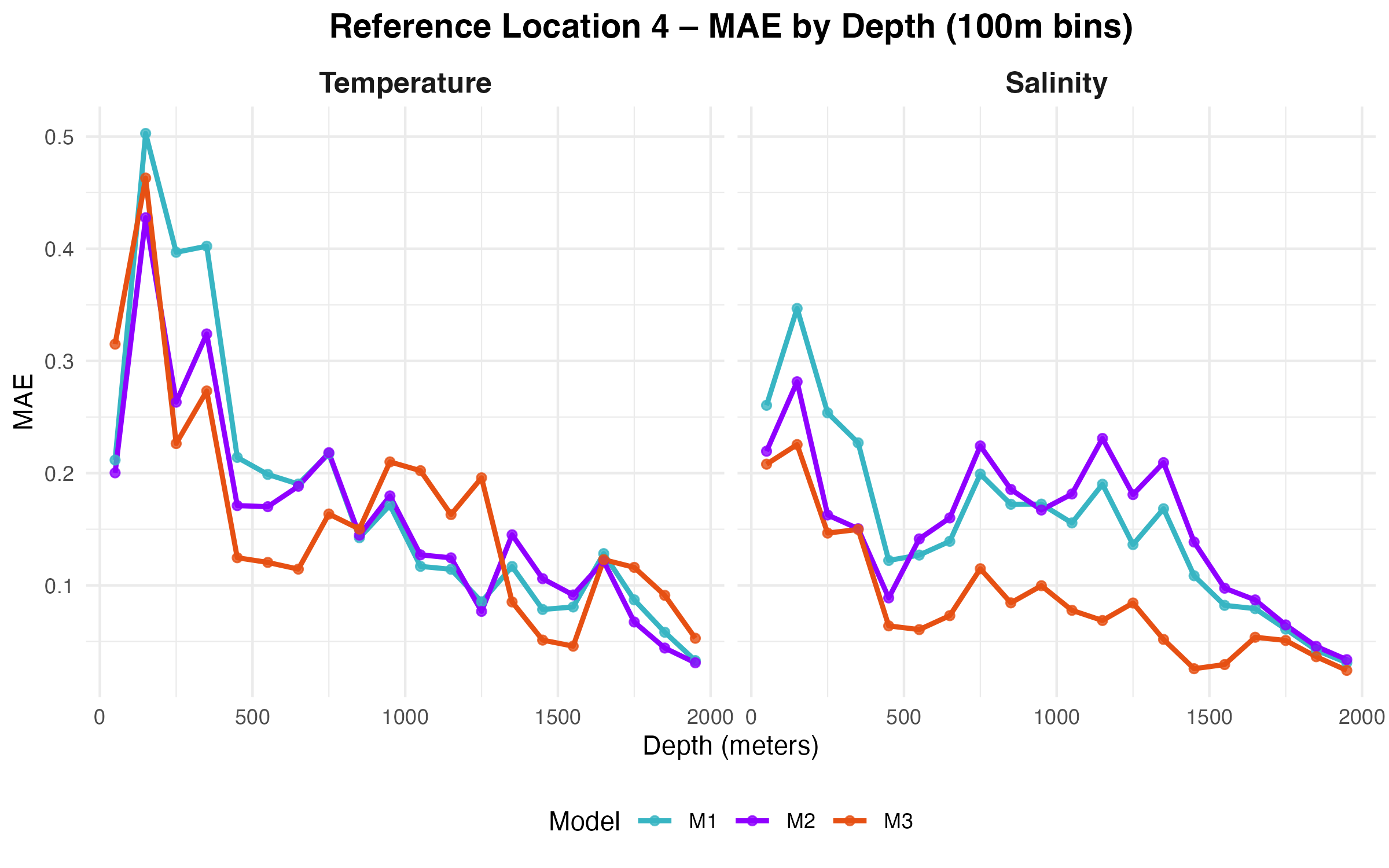}
  \end{minipage}

  \vspace{1em}
  \begin{minipage}[t]{0.45\textwidth}
    \centering
    \includegraphics[width=\textwidth]{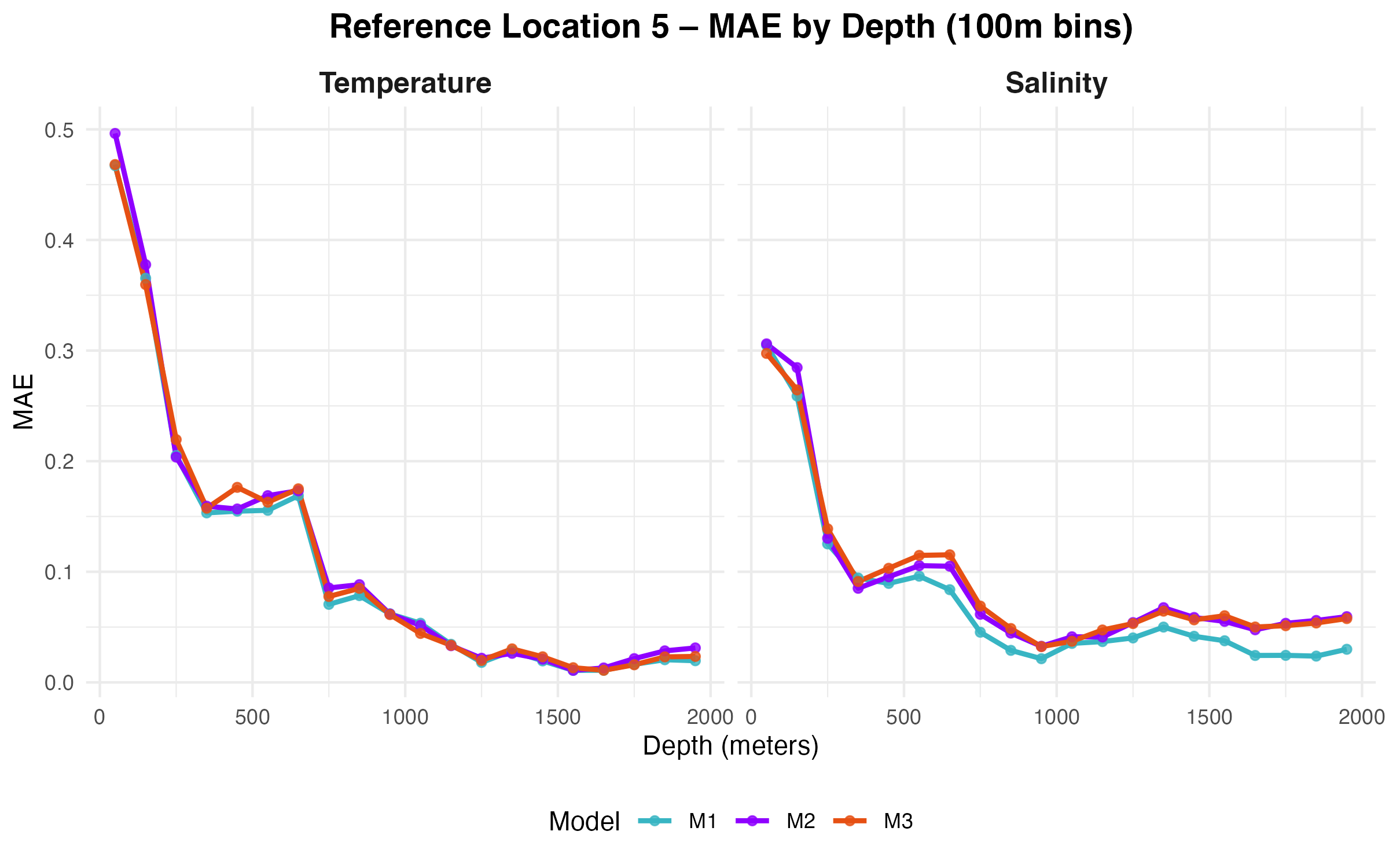}
  \end{minipage}%
  \hfill
  \begin{minipage}[t]{0.45\textwidth}
    \centering
    \includegraphics[width=\textwidth]{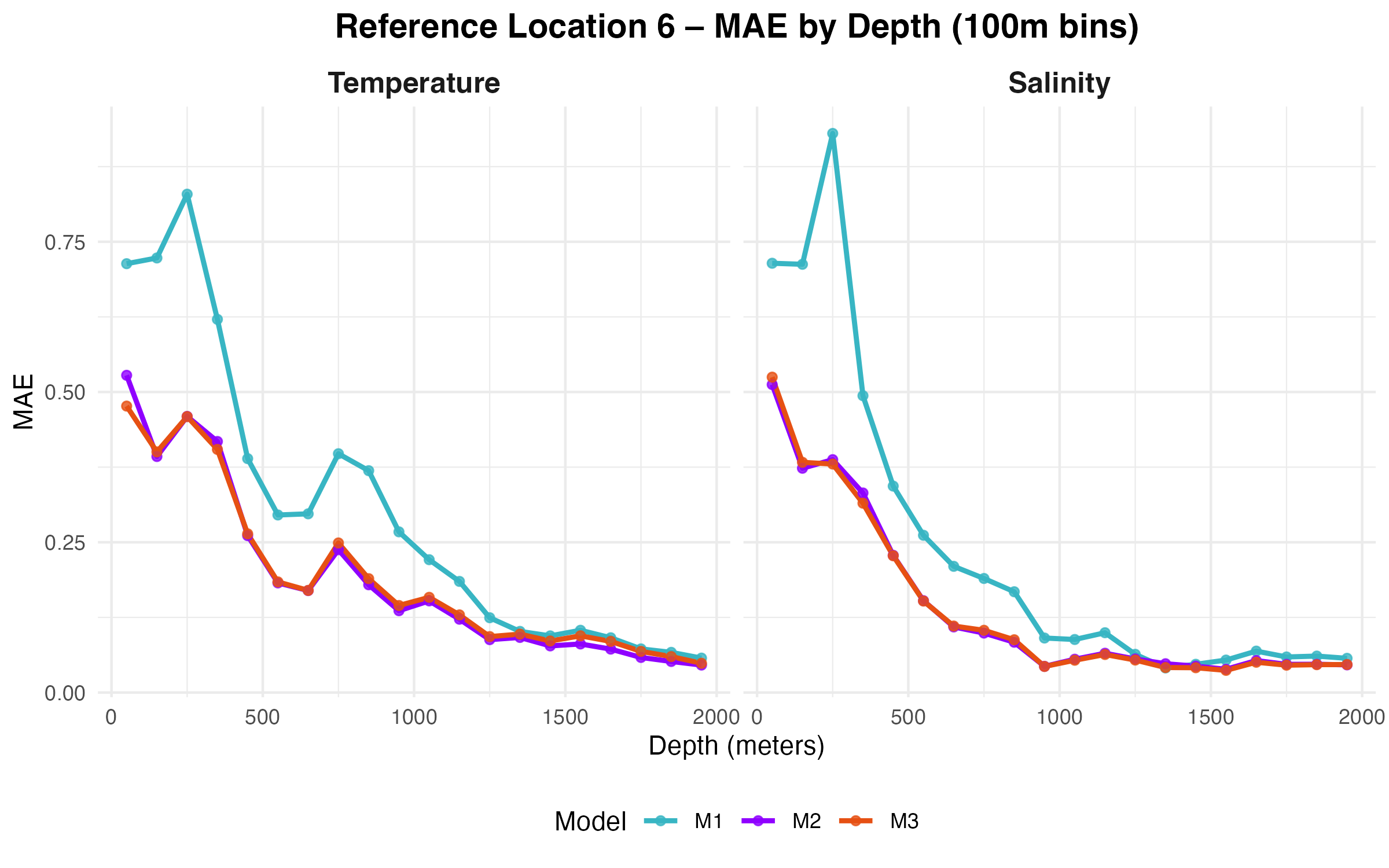}
  \end{minipage}

  \caption{MAE by depth across reference locations 1–6. Each panel shows model performance at predicting temperature and salinity as a function of depth (binned in 100 m intervals).}
  \label{fig:mae_depth}
\end{figure}

Figure~\ref{fig:mae_depth} displays the MAE as a function of depth for all six RLs. In all six locations, we observe a consistent pattern: MAE is largest in the upper 500 meters, where temperature and salinity gradients are strongest, and decreases sharply with depth. Below about 500 meters, MAE stabilizes as vertical variability diminishes.
Moreover, across all six RLs, M1 (BiMat\'ern) produces the largest kriging errors across nearly all locations—especially at RL6—reflecting its inability to represent depth-varying dependence. M2 (DiffOp--Ind) and M3 (DiffOp--Bi) show broadly similar predictive performance, but M3 exhibits a clear advantage at RL4, where temperature and salinity are nearly perfectly correlated. Because M2 assumes independence between variables, it performs poorly in this setting, even underperforming M1. This result underscores the importance of explicitly modelling cross-correlation when it is strong.

\begin{figure}[tb!]
  \centering
  \includegraphics[width=0.6\textwidth]{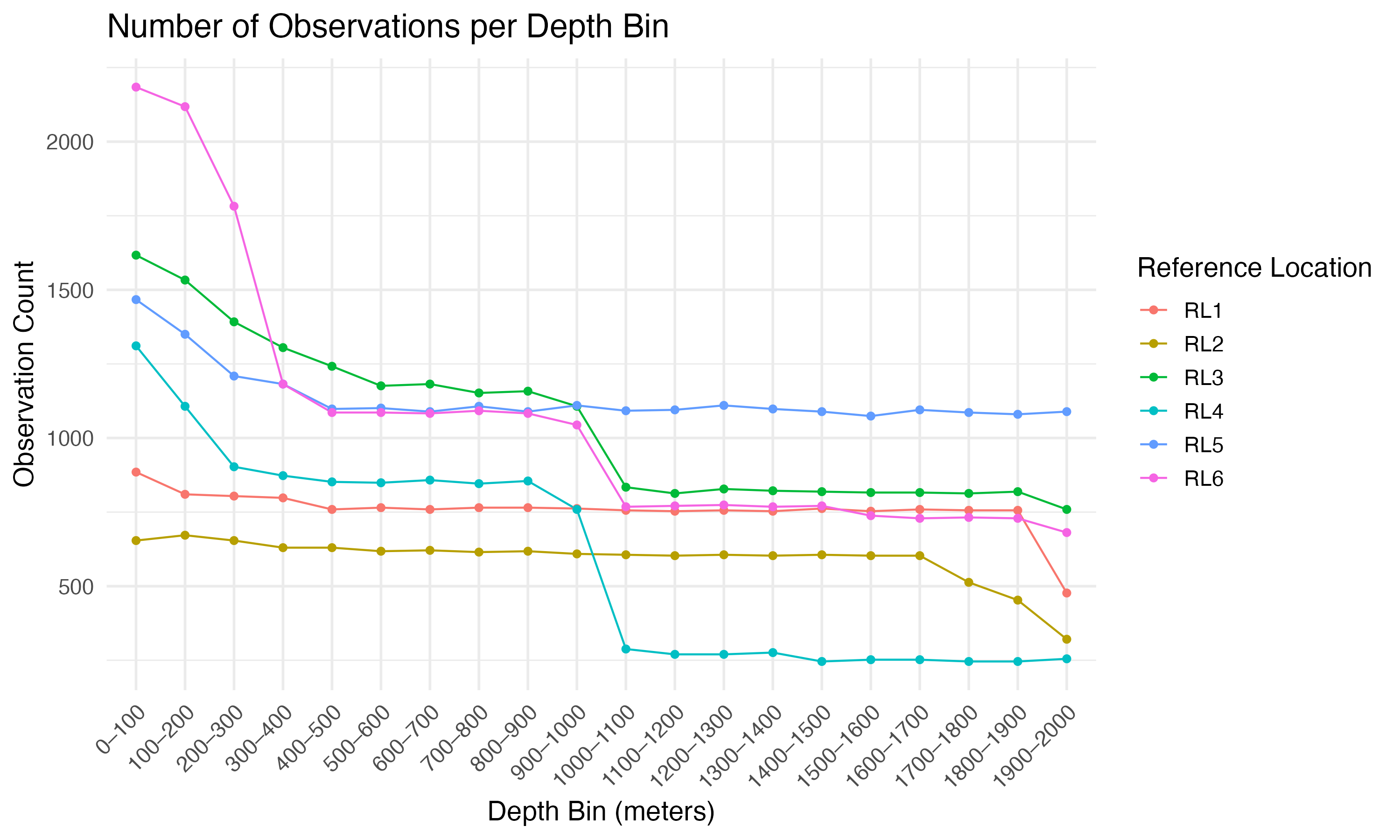}
  \caption{Number of testing observations per 100-meter depth bin for each reference location. The depth bins range from 0 to 2,000 meters, and the $x$-axis labels indicate bin boundaries.
  }
  \label{fig:obs_counts_depth}
\end{figure}

Figure~\ref{fig:obs_counts_depth} summarizes the number of available testing observations in each bin, grouped by RL. Each curve represents the bin-wise observation count for a given RL, enabling a direct comparison of vertical data coverage. This highlights substantial variability in sampling density across regions. Although the number of observations generally decreases with depth, the rate and pattern of decline differ by location, reflecting regional differences in float availability and depth penetration. Such variability in data coverage is important to consider when interpreting depth-dependent MAE patterns.

Table~\ref{tab:overall_summary} summarizes model performance across RLs using the continuous ranked probability score (CRPS) in addition to point-prediction metrics (MAE and RMSE). Models M2 and M3, both of which incorporate vertical nonstationarity, substantially outperform the stationary Mat\'ern model M1, confirming the importance of flexible vertical covariance structure for accurate inference. Moreover, M3 (DiffOp-Bi) achieves better CRPS than M2 (DiffOp-Ind) in terms of both temperature and salinity even though M2 (DiffOp-Ind) achieves slightly better point prediction in temperature, underscoring the benefits of modelling the cross-correlation for uncertainty quantification. A breakdown of the overall metrics across the six reference locations is provided in the Supplementary Material.

\begin{table}[htb!]
\centering
{
\caption{Averaged performance metrics across six reference locations (the sum of the averaged metrics across six reference locations divided by six). The subscripts $t$ and $s$ denote the temperature and salinity variables, respectively. The best model for each metric is in bold.}
\begin{tabular}{l|r|r|r}
\hline
Metric & M1 (BiMat\'ern) & M2 (DiffOp-Ind) & M3 (DiffOp-Bi)\\
\hline
$\mbox{MAE}_t$ & 0.210 & \textbf{0.167} & 0.168\\
\hline
$\mbox{MAE}_s$ & 0.037 & 0.030 & \textbf{0.028}\\
\hline
$\mbox{RMSE}_t$ & 0.372 & \textbf{0.305} & 0.310\\
\hline
$\mbox{RMSE}_s$ & 0.067 & 0.053 & \textbf{0.051}\\
\hline
$\mbox{CRPS}_t$ & 0.188 & 0.178 & \textbf{0.171}\\
\hline
$\mbox{CRPS}_s$ & 0.166 & 0.167 & \textbf{0.143}\\
\hline
\end{tabular}
\label{tab:overall_summary}
}
\end{table}

\subsection{Model diagnostics on differenced processes}

To further assess the proposed models' ability to capture spatial variation and correlation structure of the processes, we compare the empirical and fitted covariance structures of differenced processes. Specifically, we consider the first and second differencing of processes along the vertical dimension. That is, for each $Z_i(L,l,p)$, we define its first differenced process: $$Z_i^{(1)}(L,l,p)=Z_i(L,l,p) - Z_i(L,l,p+\Delta p),$$
and its second differenced process: 
$$Z_i^{(2)}(L,l,p)=-Z_i(L,l,p+\Delta p) + 2~Z_i(L,l,p)-Z_i(L,l,p-\Delta p),$$ 
with $\Delta p = 2$ meters. Differencing is performed separately for each profile and the corresponding differencing matrices are defined in Equation~\eqref{diff}.

The marginal and cross-covariance structures of the differenced bivariate process $\mathbf{Z}^{(r)}=(Z_1^{(r)}, Z_2^{(r)})^{\top}$, for $r=1,2$, are computed as linear combinations of the base covariances $C_{ij}$ from Equation~\eqref{eqn:proposed_cross_cov_explicit}. For instance, under the assumption of axial symmetry (i.e., stationarity in longitude), the first-differenced covariance is given by: 
\begin{align}
\mbox{Cov}\{Z^{(1)}_i(L,l,p), Z^{(1)}_j(L,l,p)\}  = &C_{ij}(L,L,0,p ,p)+C_{ij}(L,L,0,p+\Delta p,p+\Delta p)\nonumber \\&-C_{ij}(L,L,0,p,p+\Delta p)-C_{ij}(L,L,0,p+\Delta p,p).
\label{first-diff}
\end{align}

To carry out differencing in practice, we apply the following matrices to the data vector of each profile (ordered by increasing depth) for the first differencing ($T_1$) and second differencing ($T_2$):
\begin{align}\label{diff} T_1 =\begin{pmatrix}
1&0&0&0&\cdots&0\\
-1 & 1 & 0 &0& \cdots&0\\
0 & -1 & 1 & 0 & \cdots& 0\\
&&\ddots&\ddots&&&\\
0&0&\cdots&-1&1&0\\
0&0&\cdots&0&-1&1\\
\end{pmatrix} \mbox{   and   }~~ T_2 =\begin{pmatrix}
2&-1&0&0&\cdots&0\\
-1 & 2 & -1 &0& \cdots&0\\
0 & -1 & 2 & -1 & \cdots& 0\\
&&\ddots&\ddots&\ddots&&\\
0&0&\cdots&-1&2&-1\\
0&0&\cdots&0&-1&2\\
\end{pmatrix}. \end{align}
If $\Sigma=(\Sigma_{ij})_{i,j=1,2}$ denotes the $2\times 2$ block covariance matrix of temperature and salinity residuals for a given profile, then the covariance matrix of the differenced process is: \begin{align}
\Sigma^{(r)}=T_r \Sigma_{ij} T_r^T,
\label{diff_cov}
\end{align} 
for $r=1$ (first differencing) or $r=2$ (second differencing). 

To obtain empirical covariance values from the Argo data, we apply first or second differencing along the vertical pressure levels for each float. This is justified by the fact that most observations within a single float profile share the same longitude and latitude. Furthermore, between depths of 0 and approximately 1,200 meters, the pressure levels are typically sampled at 2-meter intervals, while sampling becomes sparser below 1,200 meters. For each level $p_m$, we calculate the difference between temperature (or salinity) residuals at the level $p_m$ and the adjacent pressure levels, $p_m+\Delta p$ and $p_m - \Delta p$ for each float with $\Delta p =$ 2 meters, except the very top and very bottom levels, where differencing follows the scheme in Equation~\eqref{diff}. Empirical variances and covariances are then computed using the differenced values from all floats within the local region. Since the spatial domain is narrow, we reasonably assume stationarity within each local region when aggregating these empirical quantities.

\begin{figure}[tb!]
	\centering
\includegraphics[scale=0.65]{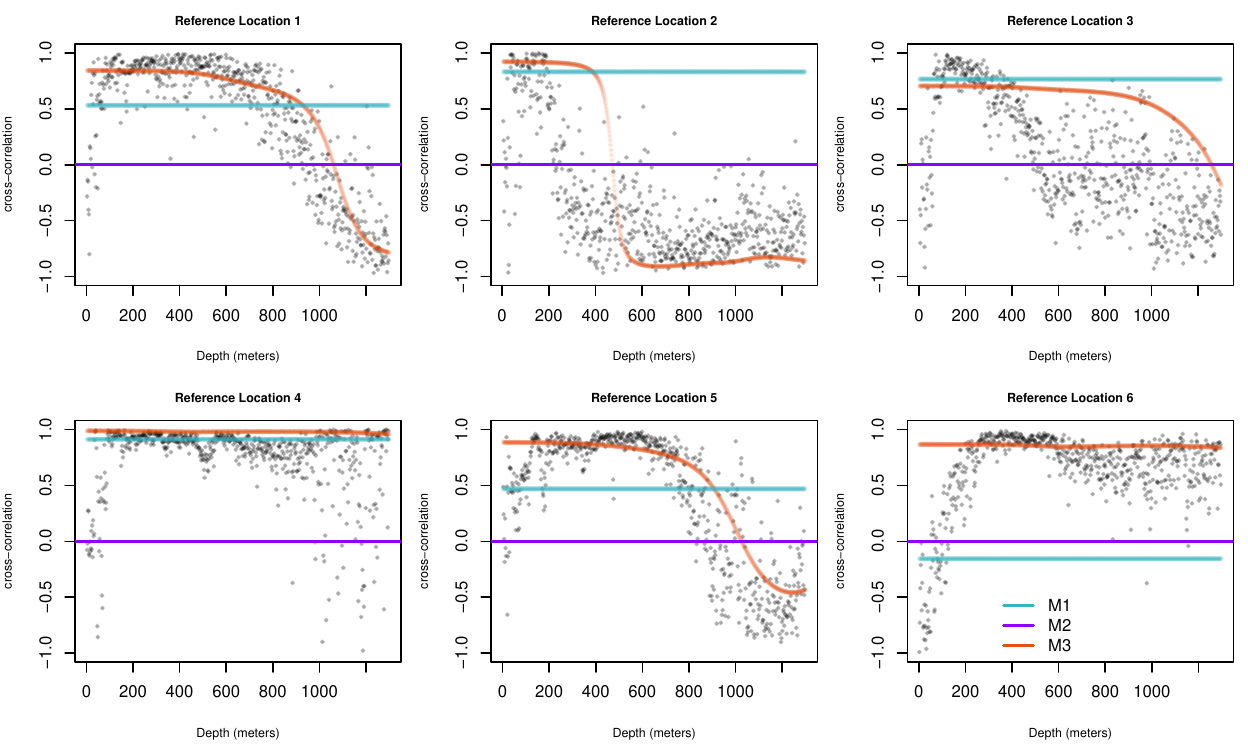} 
     \caption{Empirical (black) and model-based estimates of colocated correlations of the first differenced processes, plotted as a function of depth at each reference location.}
     \label{first-difference}
\end{figure}

\begin{figure}[tb!]
	\centering
		\includegraphics[scale=0.65]{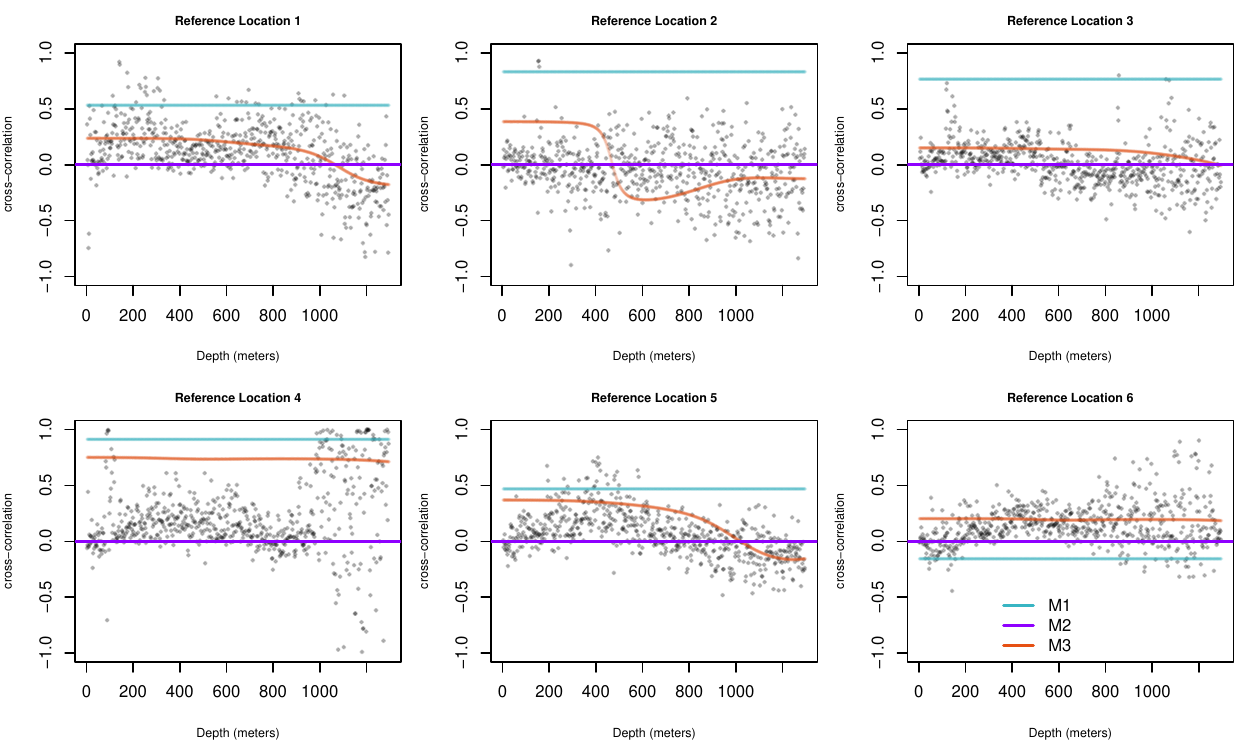} 
     \caption{Empirical (black) and model-based estimates of colocated correlations of the second differenced processes, plotted as a function of depth at each reference location.}
     \label{second-difference}
\end{figure}

Figures~\ref{first-difference} and~\ref{second-difference} present model diagnostics based on the first- and second-differenced processes, respectively, across all six RLs. At each depth level, a single empirical value is computed by aggregating (differenced) observations across all Argo floats within a local region, making use of the assumption of local stationarity. Each bin represents a depth interval of 2 meters.
The plotted model estimates (M1 (BiMat\'ern), M2 (DiffOp-Ind), M3 (DiffOp-Bi)) appear as smooth curves per model, consistent with the assumption of horizontal stationarity within each region. M3 (DiffOp-Bi)—our proposed nonstationary cross-covariance model—consistently aligns closely with the empirical estimates in both the first and second-differenced diagnostics. This strong agreement confirms that M3 (DiffOp-Bi) not only captures the colocated correlation on the original scale but also preserves realistic vertical gradients and curvature in the data. By contrast, M1 (BiMat\'ern) assumes a constant cross-correlation and produces flat lines across depth, while M2 (DiffOp-Ind) assumes independence and yields zero correlation throughout. These assumptions visibly misrepresent the empirical behavior, particularly in the second-differenced case. M1 (BiMat\'ern) substantially underestimates the magnitude and variation of the empirical curves, most notably at RLs 2, 3, and 5. Even in the first-differenced case, M1 (BiMat\'ern) fails to track the observed structure except in regions where the empirical curve happens to be flat. Notably, at RL4, none of the models match the empirical second-difference curve well. This suggests a more complex dependence structure at that site that may not be fully captured by any of the current models and warrants further investigation. Together, these diagnostics validate the robustness of M3 (DiffOp-Bi) in reproducing depth-varying correlation structures, even after applying differencing operators that amplify local curvature. This capacity to preserve second-order structure under transformation is a strong indication of M3 (DiffOp-Bi)’s fidelity to the true data-generating process.

\section{Discussion} \label{sec:discussion}

Argo floats play a critical role in sampling temperature and salinity throughout the world’s oceans. However, despite the global reach of the Argo network, horizontal and vertical gaps remain. Many locations are undersampled, and vertical resolution is often sparse, with numerous floats not covering the full depth range. In this work, we contributed to the ongoing efforts to produce high-quality interpolations of temperature and salinity at unsampled locations through bivariate modelling of Argo residuals. Motivated by the underlying physical coupling between temperature and salinity, we extended traditional spatial modelling by incorporating vertical structure—an essential feature of ocean dynamics. Consistent with findings from previous studies, we showed that the correlation between temperature and salinity varies significantly with depth. Our proposed model captures this vertical nonstationarity using a flexible, depth-varying cross-covariance function. The resulting framework accommodates diverse oceanographic conditions and vertical regimes, offering a robust statistical approach for 3D mapping of environmental variables. Building on this foundation, future work could explore extensions such as allowing spatial range parameters to vary with depth and developing a 4D space-time model.

\bibliographystyle{apalike}
\bibliography{references}

\newpage

\appendix

\section*{Appendix}

\renewcommand{\thefigure}{A.\arabic{figure}}
\setcounter{figure}{0}  

\section{Assessment of Gaussianity Assumption}

\begin{figure}[H]
    \centering
    \begin{adjustbox}{width=\textwidth}
    \begin{tabular}{ccc}
        \includegraphics[width=0.32\textwidth]{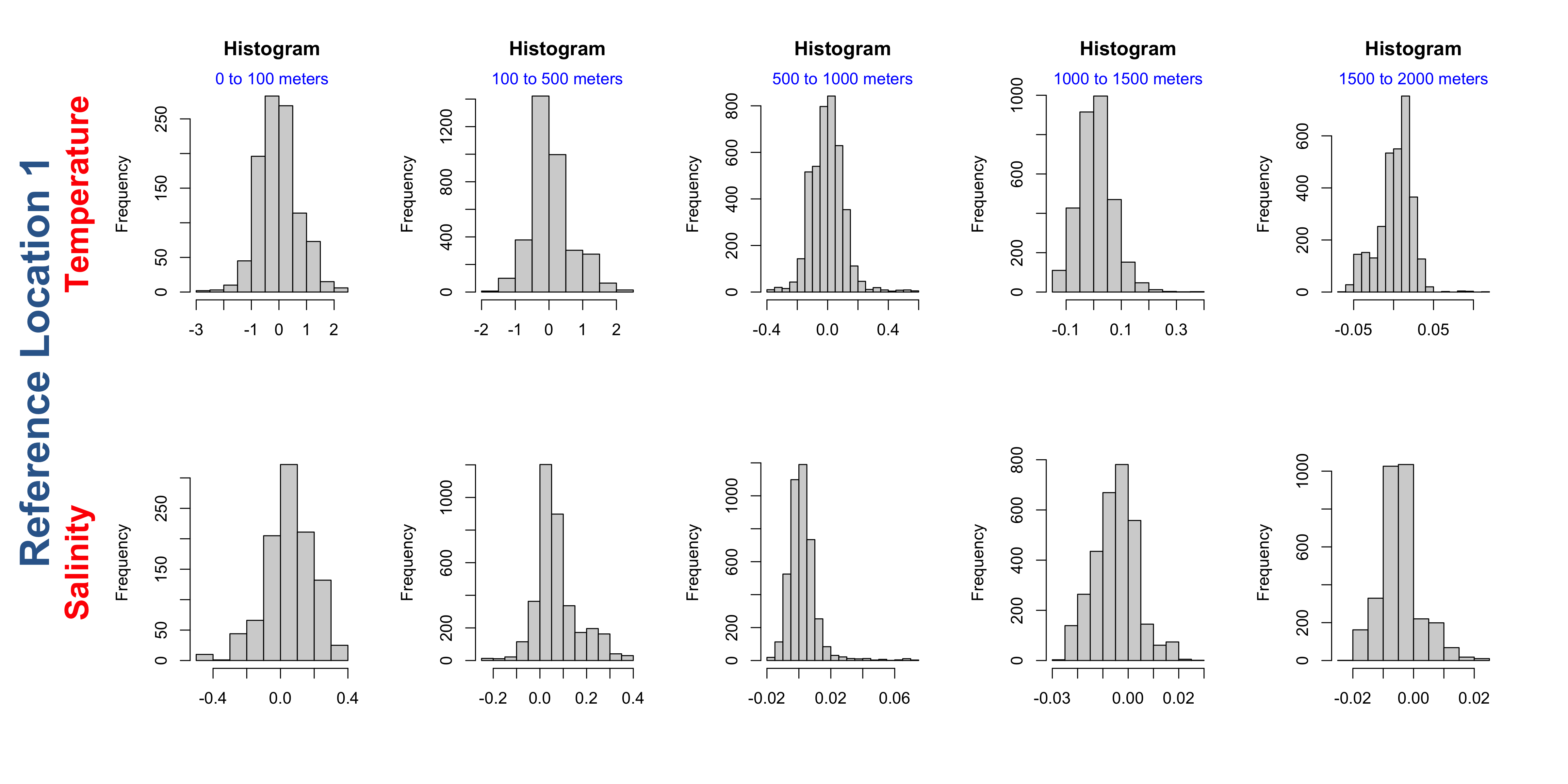} &
        \includegraphics[width=0.32\textwidth]{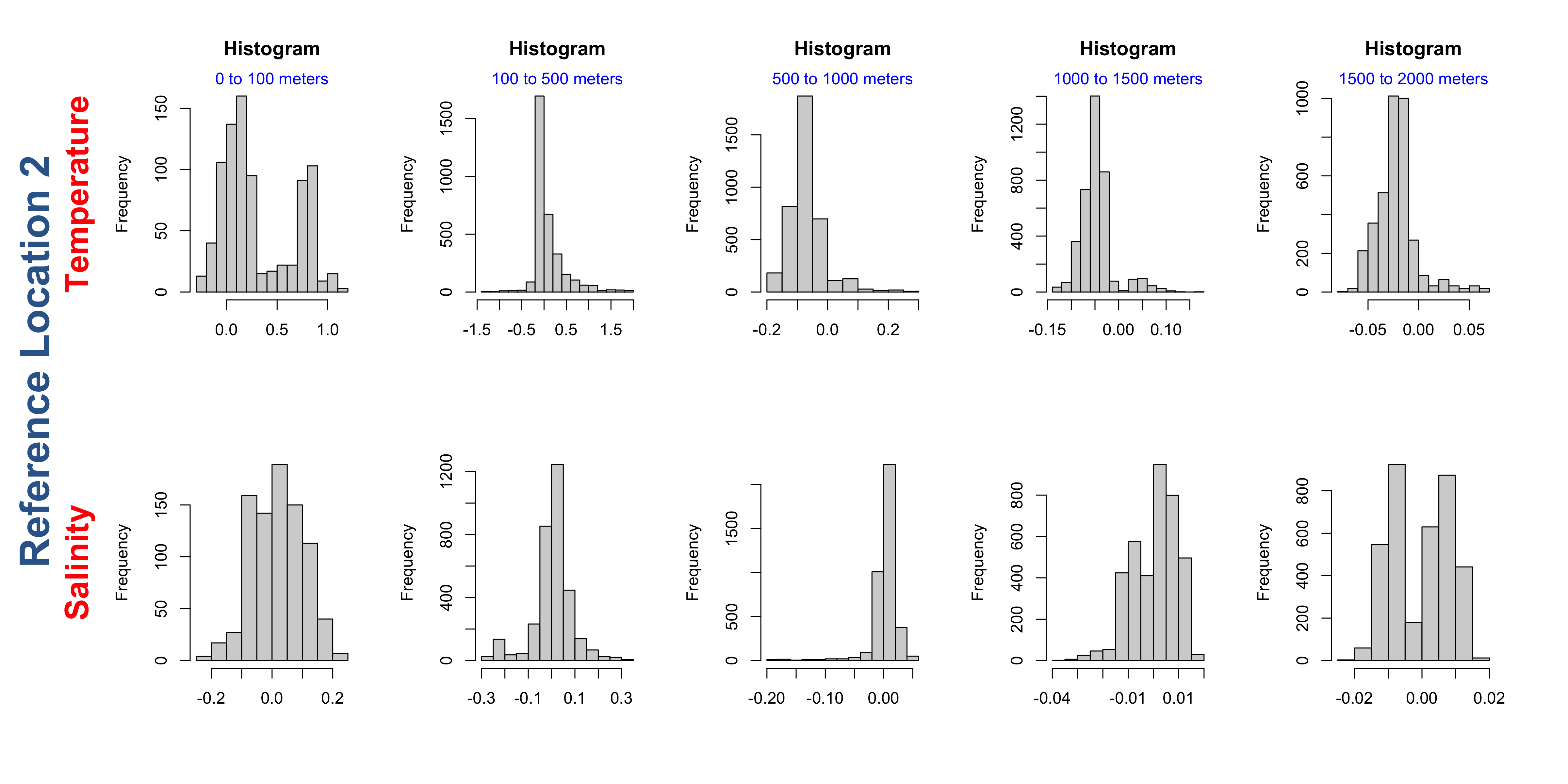} &
        \includegraphics[width=0.32\textwidth]{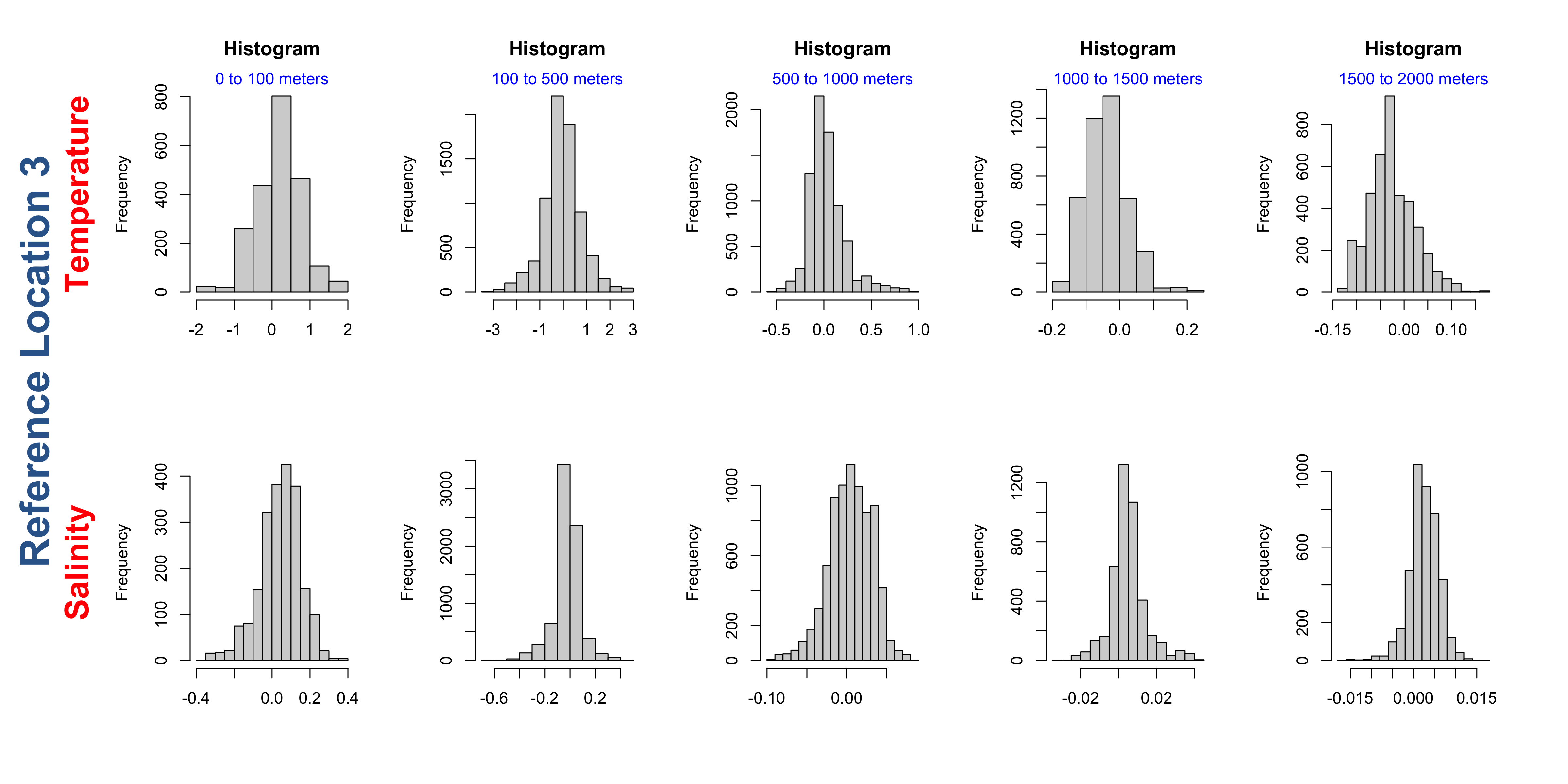} \\
        \includegraphics[width=0.32\textwidth]{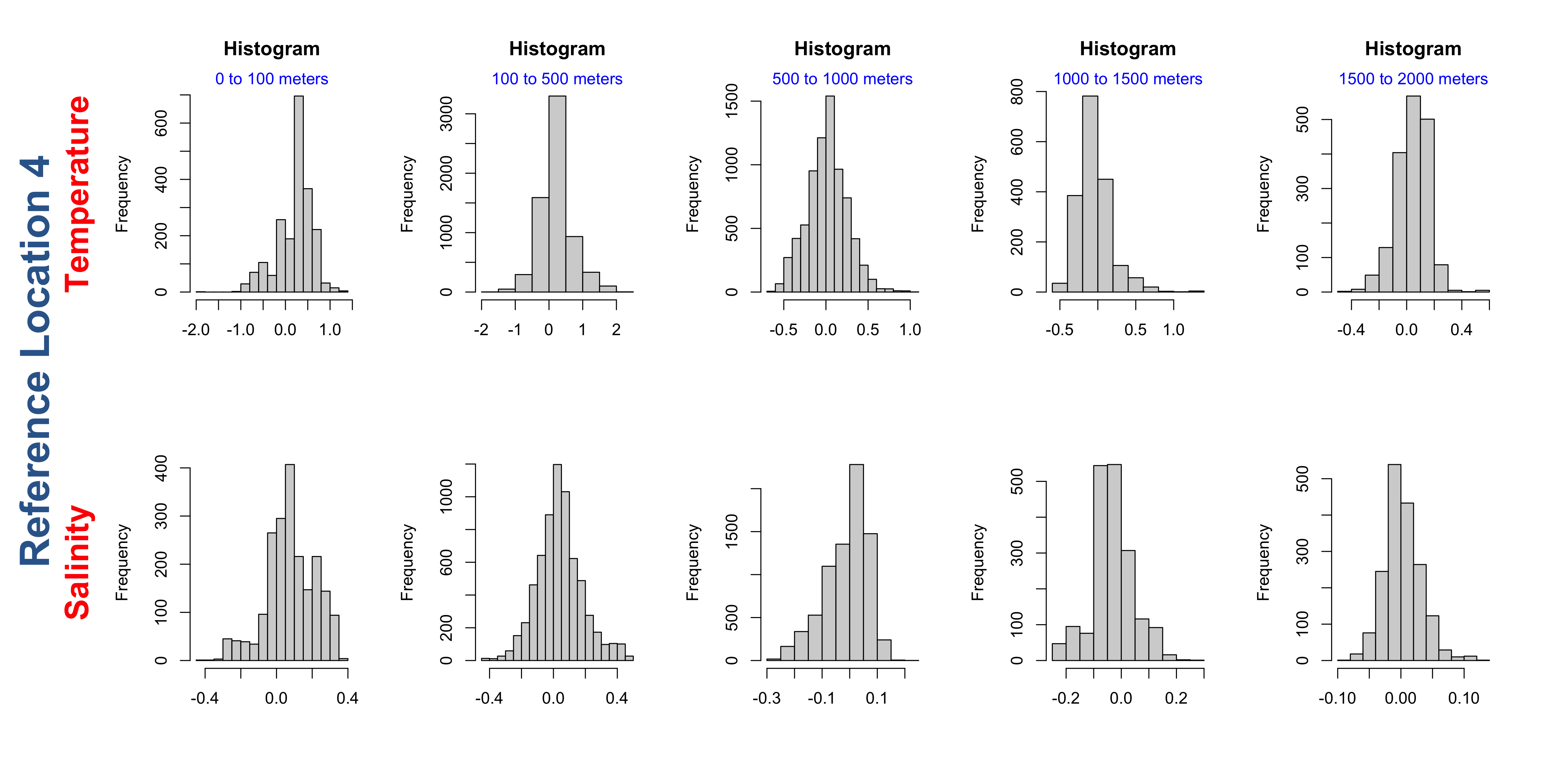} &
        \includegraphics[width=0.32\textwidth]{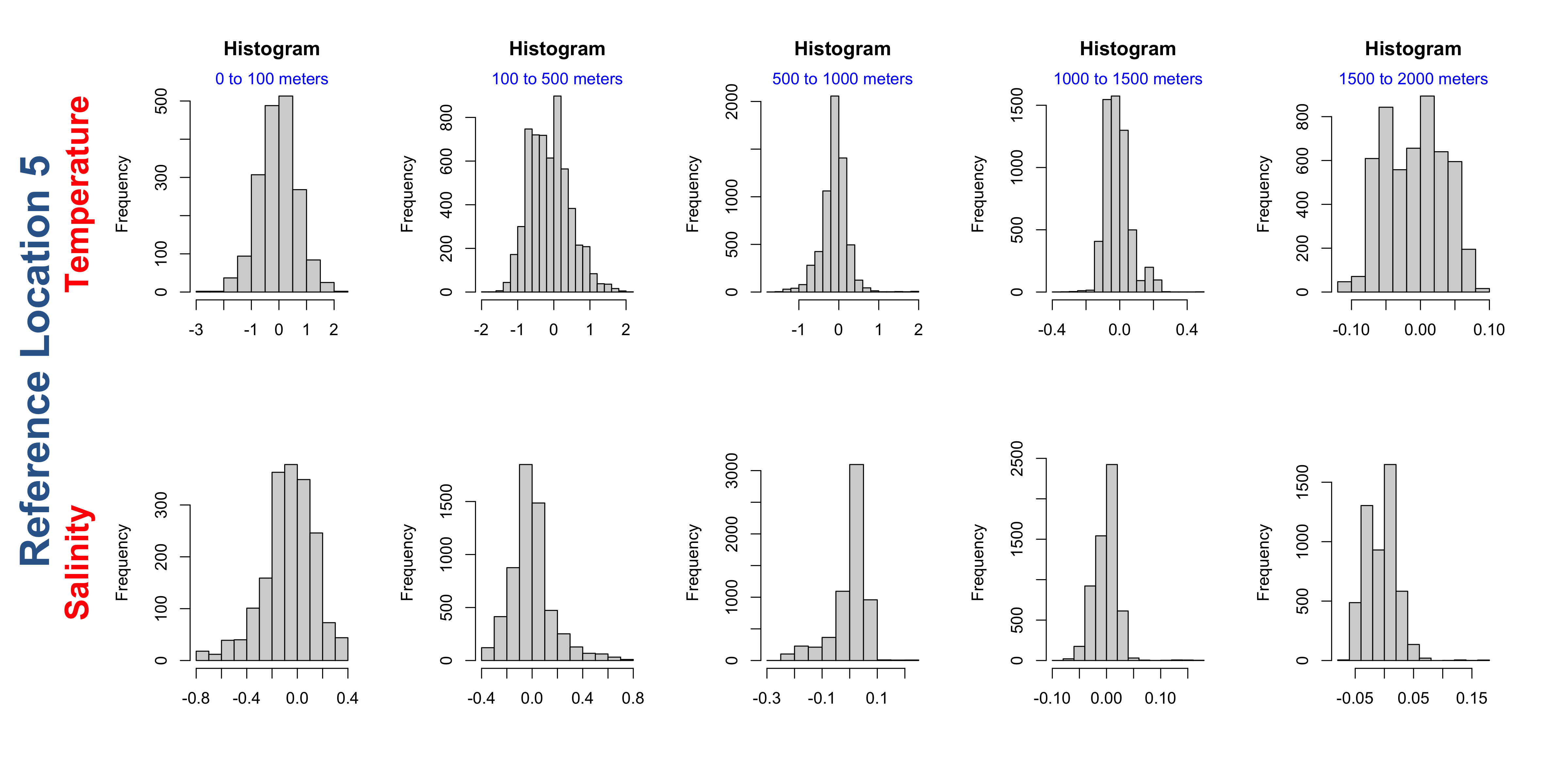} &
        \includegraphics[width=0.32\textwidth]{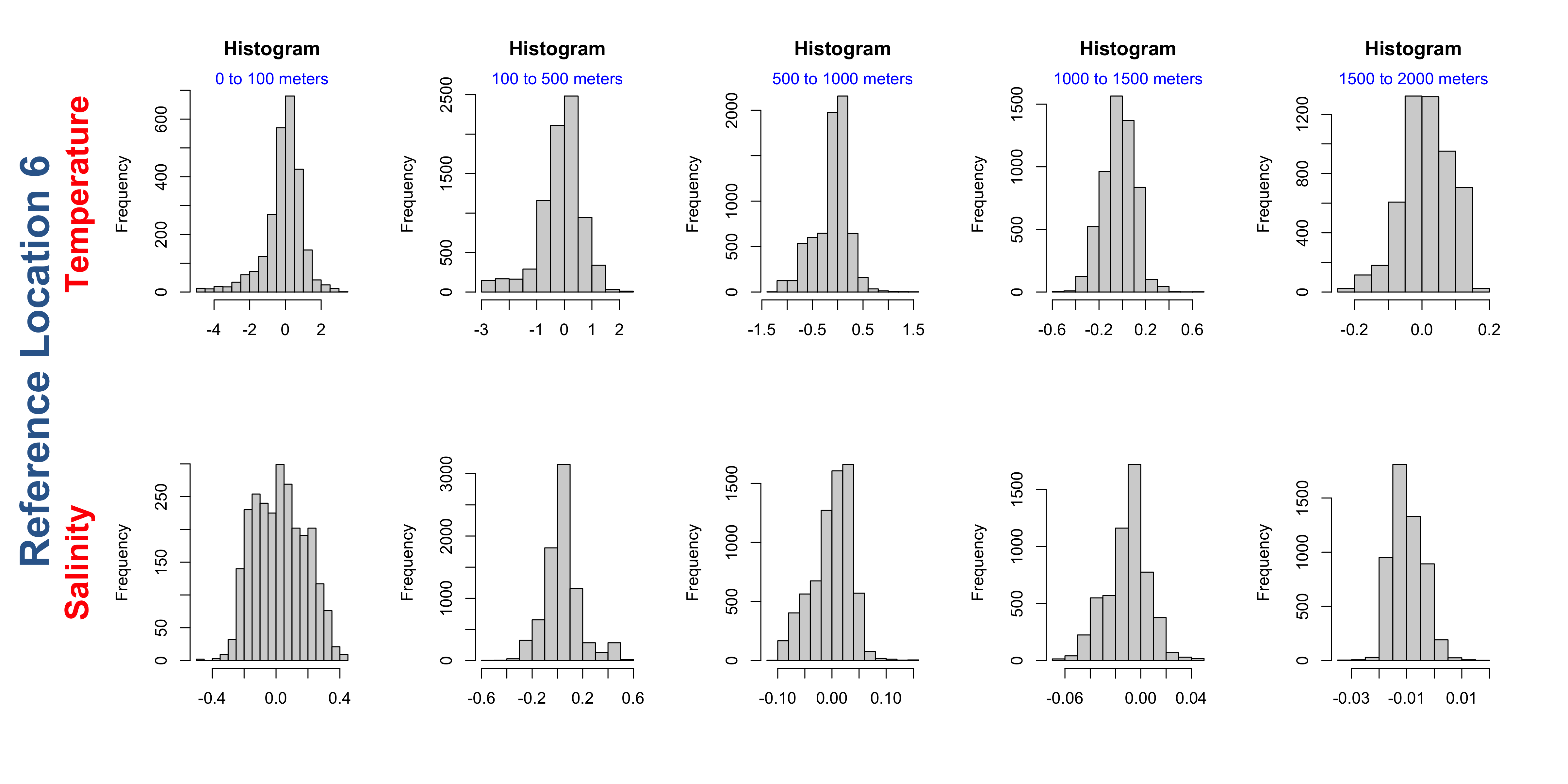}
    \end{tabular}
    \end{adjustbox}
    \caption{Histograms of standardized temperature (top row) and salinity (bottom row) residuals across five depth intervals for each of the six reference locations.}
    \label{fig:histograms}
\end{figure}

\begin{figure}[H]
    \centering
    \begin{adjustbox}{width=\textwidth}
    \begin{tabular}{ccc}
        \includegraphics[width=0.32\textwidth]{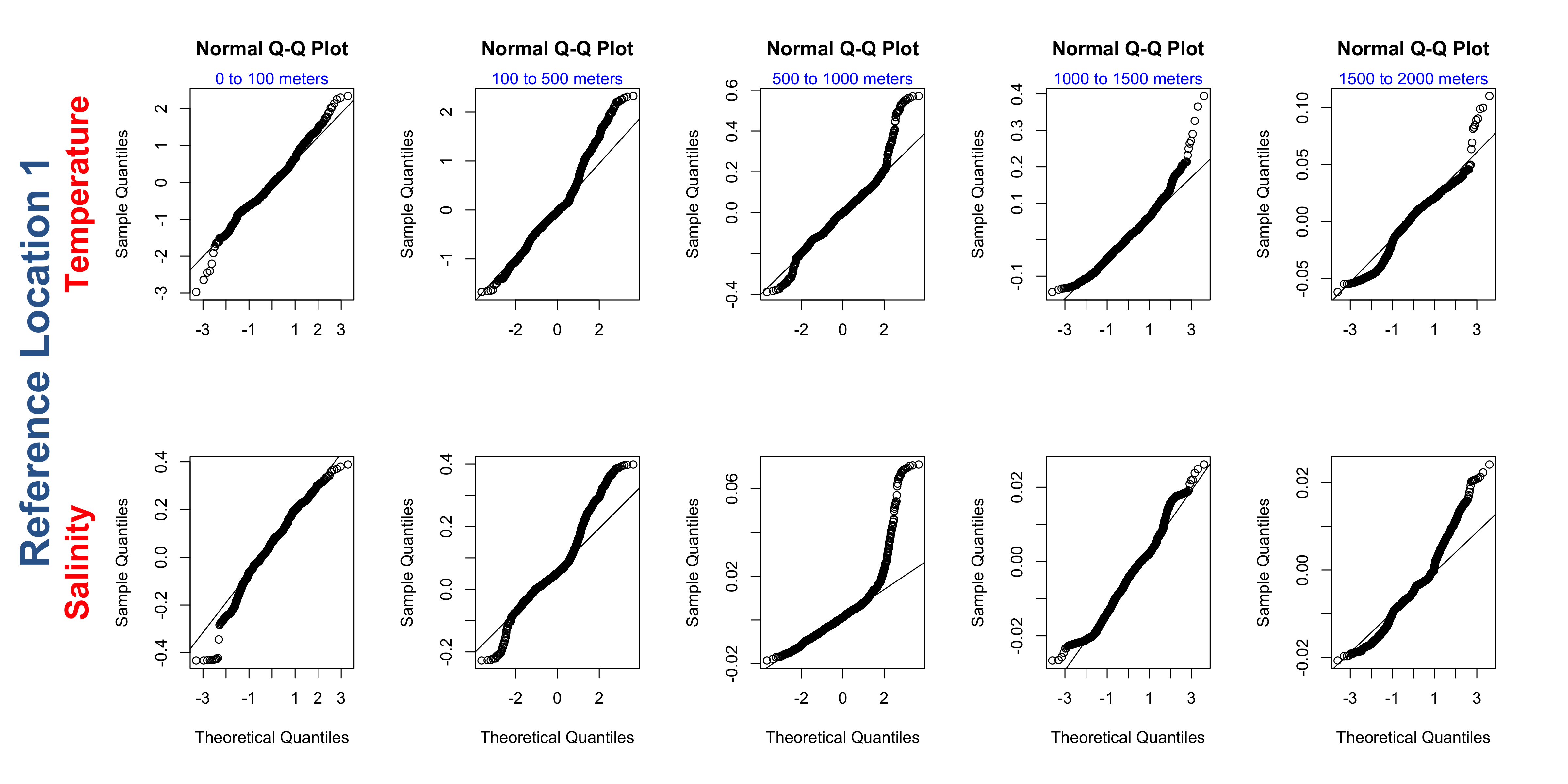} &
        \includegraphics[width=0.32\textwidth]{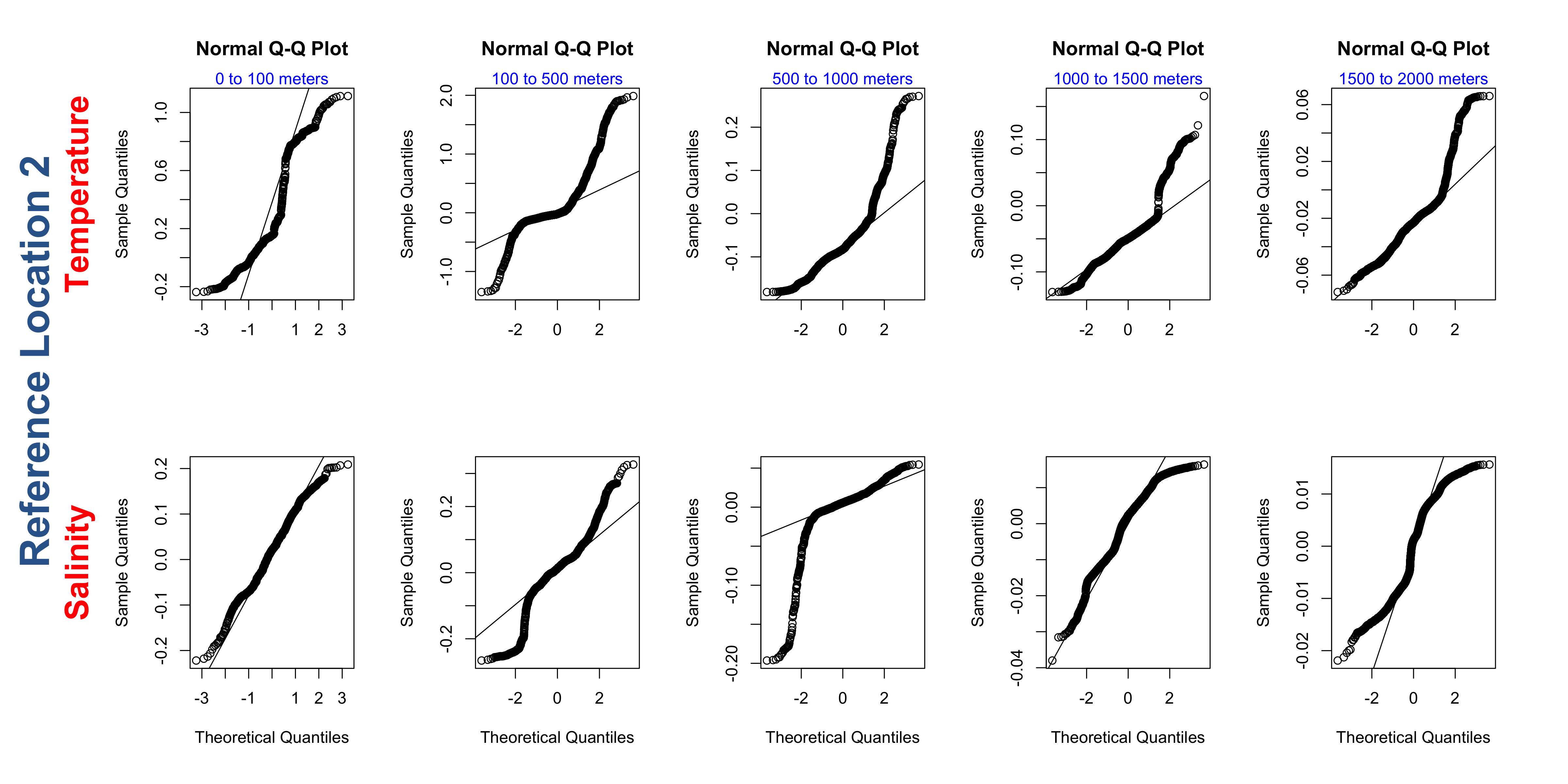} &
        \includegraphics[width=0.32\textwidth]{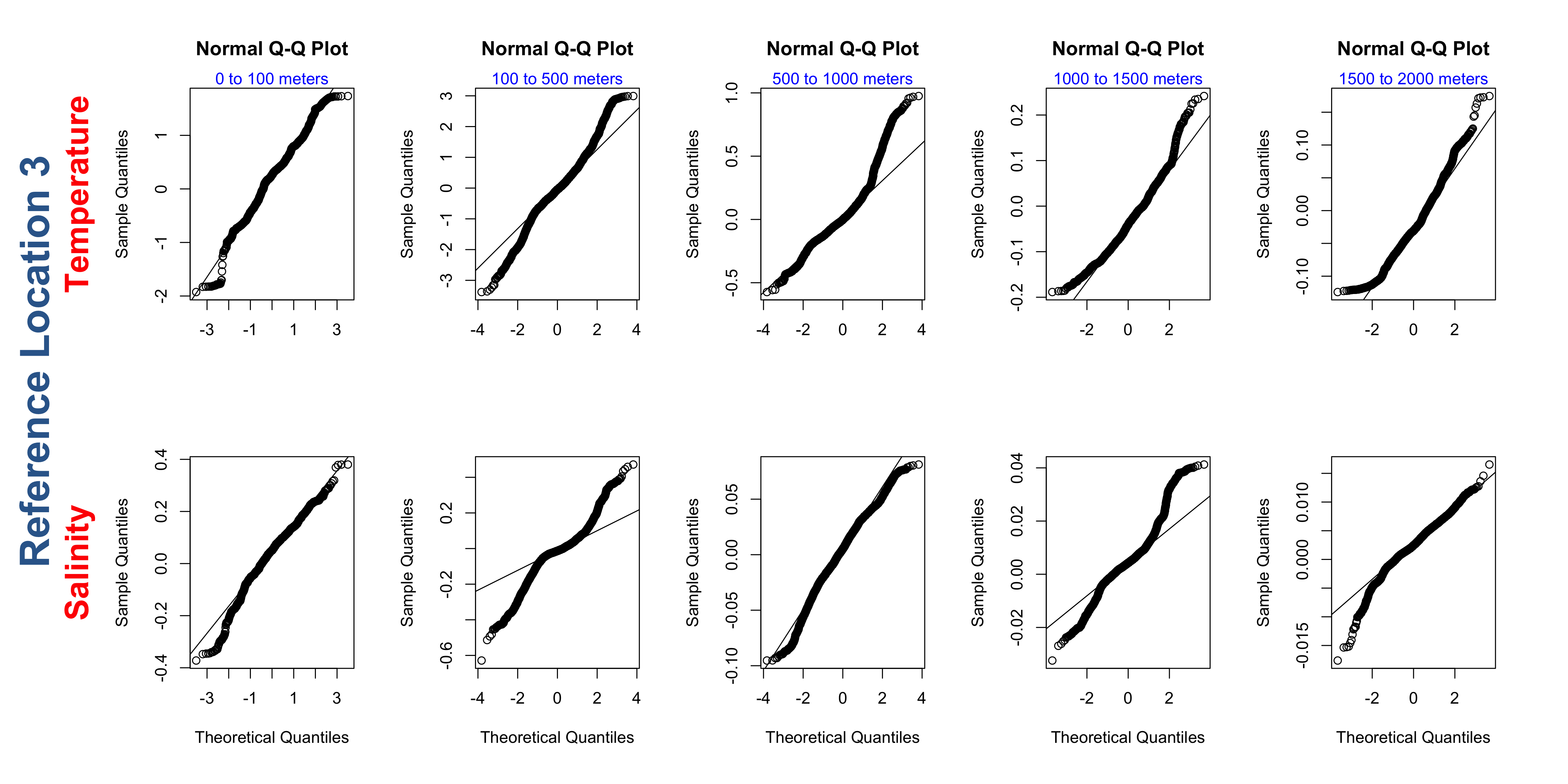} \\
        \includegraphics[width=0.32\textwidth]{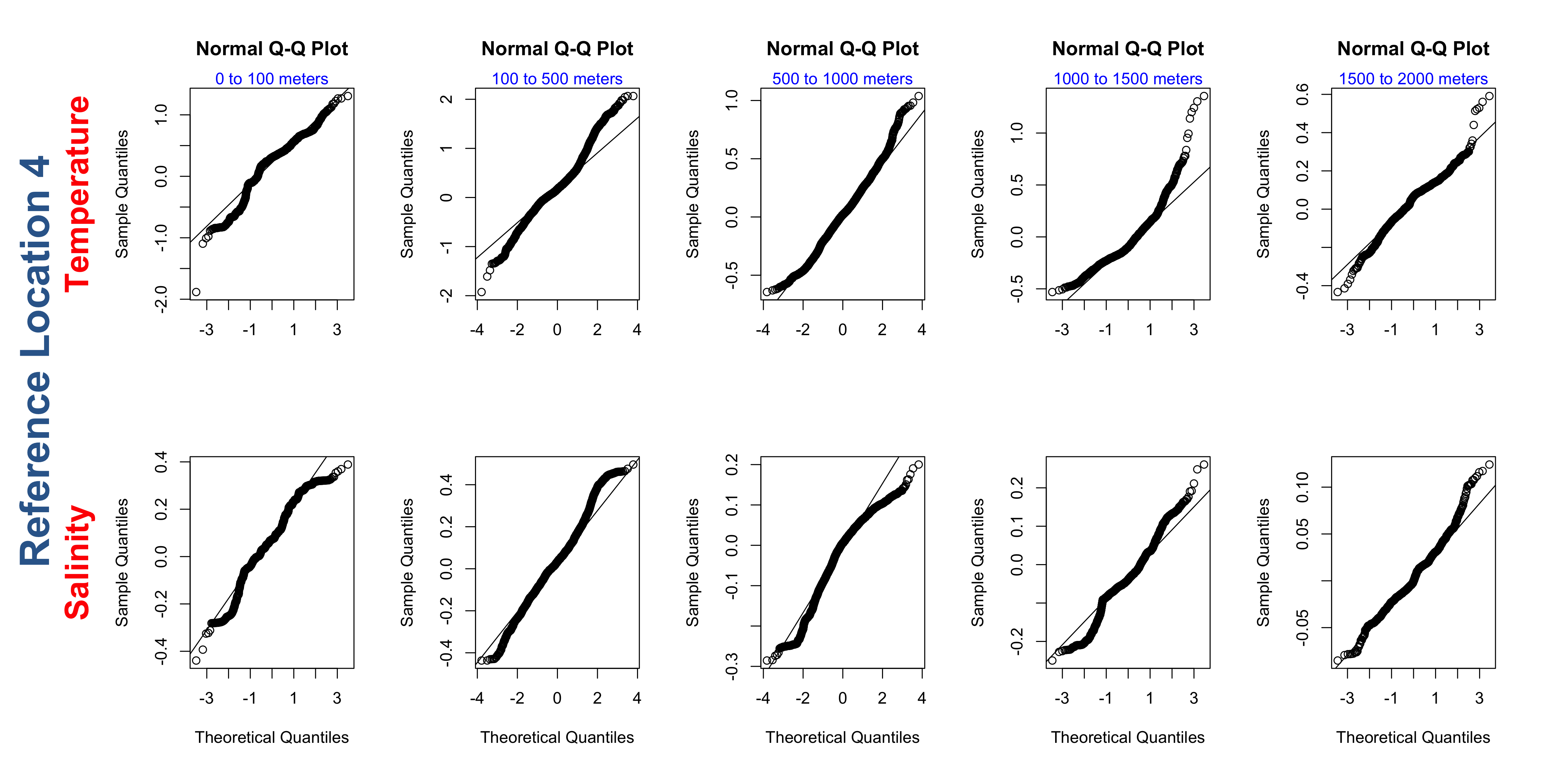} &
        \includegraphics[width=0.32\textwidth]{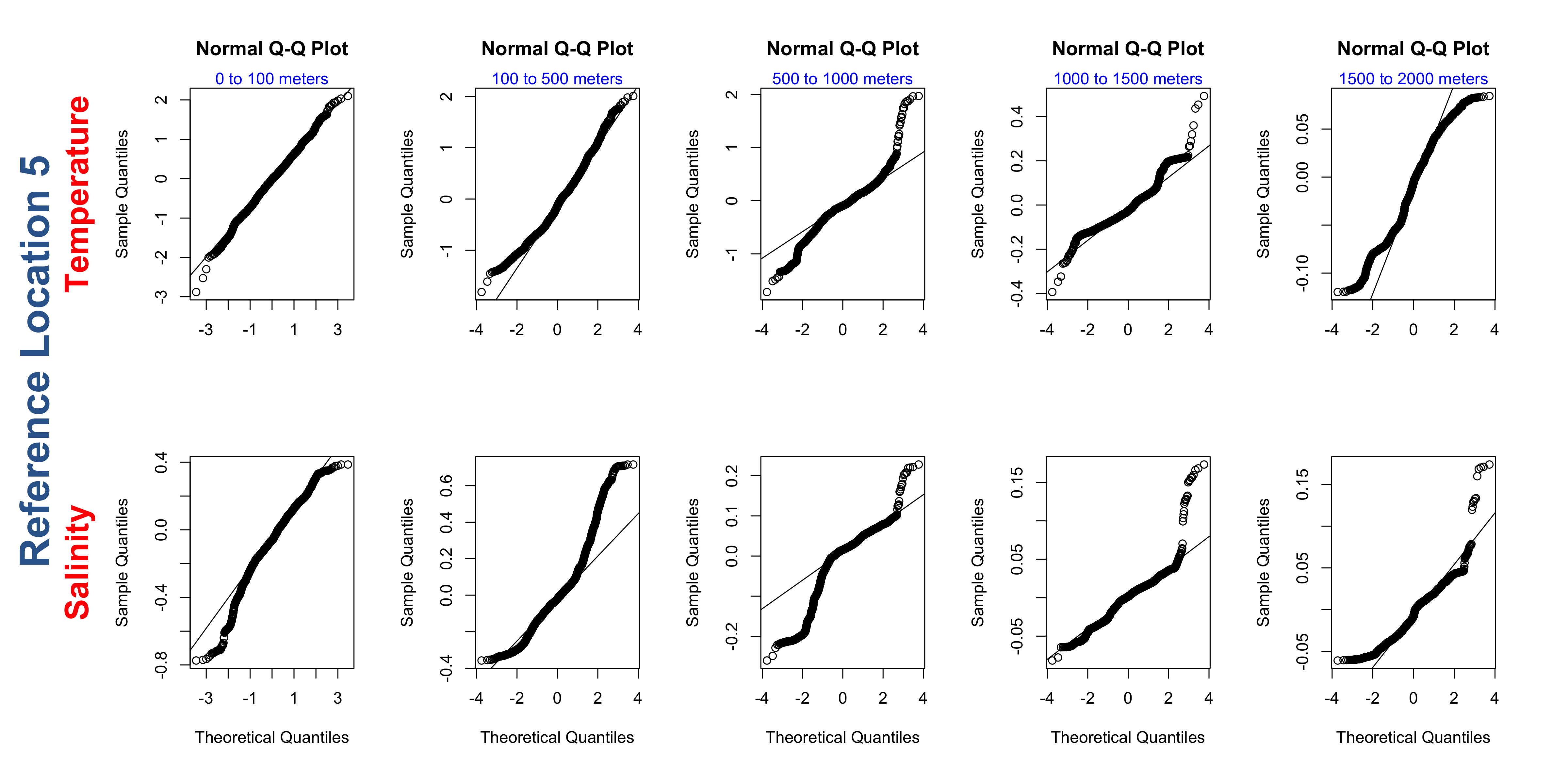} &
        \includegraphics[width=0.32\textwidth]{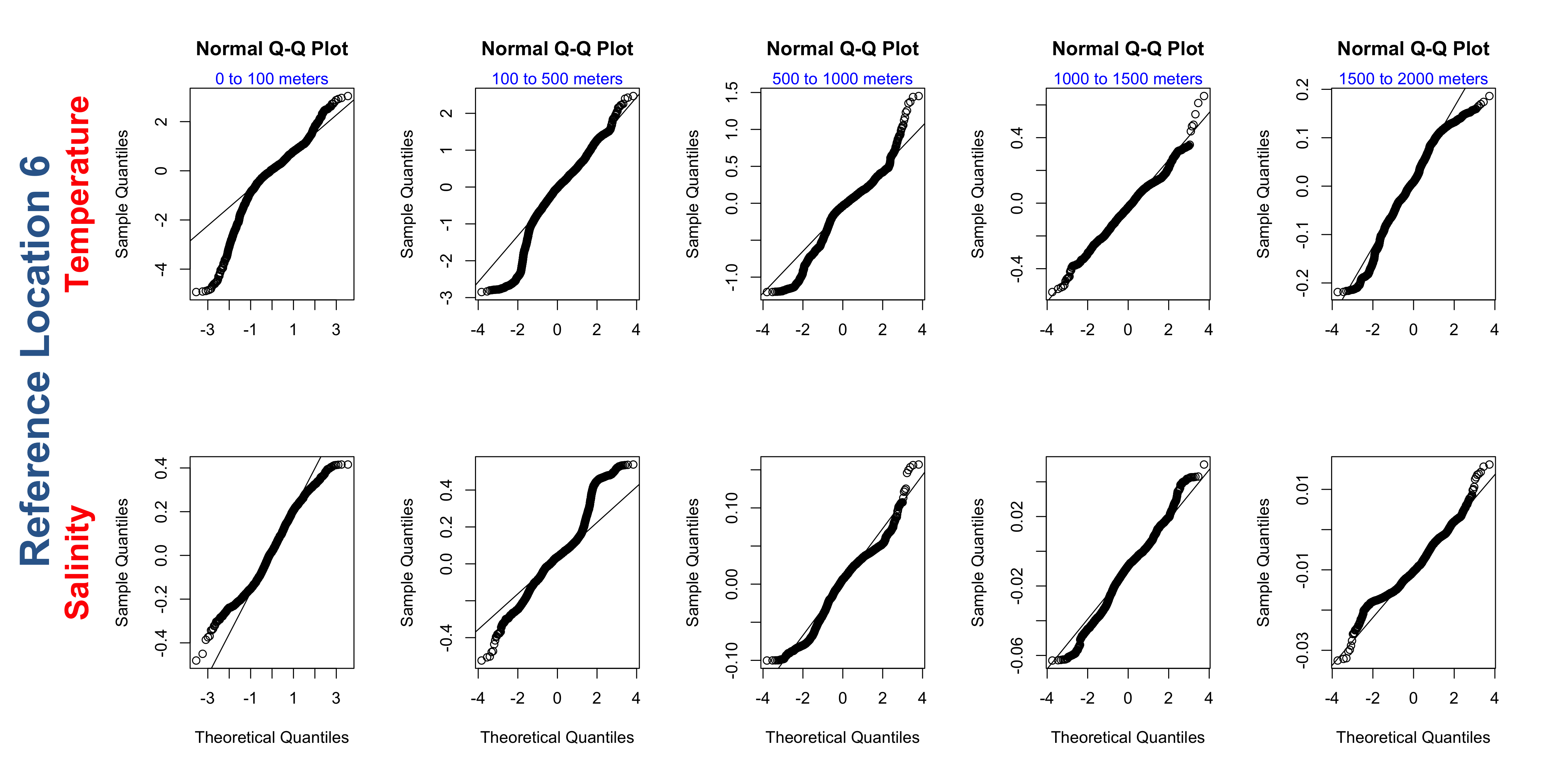}
    \end{tabular}
    \end{adjustbox}
    \caption{Q--Q plots of standardized temperature (top row) and salinity (bottom row) residuals across five depth intervals for each of the six reference locations.}
    \label{fig:qqplots}
\end{figure}

To evaluate the Gaussianity assumption, we examine the distribution of temperature and salinity residuals across six reference locations and five depth layers (0–100 m, 100–500 m, 500–1000 m, 1000–1500 m, and 1500–2000 m). Figures~\ref{fig:histograms} and~\ref{fig:qqplots} present corresponding histograms and Q–Q plots for each location and depth bin. Overall, residuals are approximately Gaussian, with better conformity in deeper layers. Shallow regions show some deviations—such as skewness or heavy tails—likely due to surface variability, stratification, or unresolved mesoscale dynamics. These deviations diminish with depth, suggesting improved homogeneity and model fit at lower layers. Importantly, none of the departures are severe enough to undermine the Gaussian modeling assumption used in kriging and inference.

\end{document}